\documentclass[twocolumn,figuresright]{aastex631}

\usepackage{multirow}
\usepackage{supertabular}
\usepackage{amsmath} 
\usepackage{amssymb}
\usepackage{makecell}
\usepackage{soul}
\usepackage{ulem}
\usepackage{rotating}[figuresright]
\usepackage{float}
\usepackage{booktabs}

\newcommand{\HXMT}{\textit{Insight}-HXMT }

\shorttitle{}
\shortauthors{Yan-Qiu Zhang et al.}
\graphicspath{{./}{figures/}}

\begin{document}


\title{Multi-Instrument Search for Gamma-Ray Counterpart of X-ray Transients detected by EP/WXT}

\correspondingauthor{Ce Cai, Shao-Lin Xiong}

\email{caice@hebtu.edu.cn, xiongsl@ihep.ac.cn}

\author{Yan-Qiu Zhang}
\affiliation{State Key Laboratory of Particle Astrophysics, Institute of High Energy Physics, Chinese Academy of Sciences, Beijing 100049, China}
\affiliation{University of Chinese Academy of Sciences, Chinese Academy of Sciences, Beijing 100049, China}

\author{Wang-Chen Xue}
\affiliation{State Key Laboratory of Particle Astrophysics, Institute of High Energy Physics, Chinese Academy of Sciences, Beijing 100049, China}
\affiliation{University of Chinese Academy of Sciences, Chinese Academy of Sciences, Beijing 100049, China}

\author{Jin-Peng Zhang}
\affiliation{State Key Laboratory of Particle Astrophysics, Institute of High Energy Physics, Chinese Academy of Sciences, Beijing 100049, China}
\affiliation{University of Chinese Academy of Sciences, Chinese Academy of Sciences, Beijing 100049, China}

\author{Ce Cai*}
\affiliation{College of Physics and Hebei Key Laboratory of Photophysics Research and Application, Hebei Normal University, Shijiazhuang, \\ Hebei 050024, China}
\affiliation{Shijiazhuang Key Laboratory of Astronomy and Space Science, Hebei Normal University, Shijiazhuang, Hebei 050024, China}

\author{Shao-Lin Xiong*}
\affiliation{State Key Laboratory of Particle Astrophysics, Institute of High Energy Physics, Chinese Academy of Sciences, Beijing 100049, China}

\author{Cheng-Kui Li}
\affil{State Key Laboratory of Particle Astrophysics, Institute of High Energy Physics, Chinese Academy of Sciences, Beijing 100049, China}

\author{Yuan Liu}
\affil{National Astronomical Observatories, Chinese Academy of Sciences, Beijing 100101, China}

\author{Chen-Wei Wang}
\affiliation{State Key Laboratory of Particle Astrophysics, Institute of High Energy Physics, Chinese Academy of Sciences, Beijing 100049, China}
\affiliation{University of Chinese Academy of Sciences, Chinese Academy of Sciences, Beijing 100049, China}

\author{Hao-Xuan Guo}
\affiliation{State Key Laboratory of Particle Astrophysics, Institute of High Energy Physics, Chinese Academy of Sciences, Beijing 100049, China}
\affiliation{Department of Nuclear Science and Technology, School of Energy and Power Engineering, Xi'an Jiaotong University, Xi'an, China}

\author{Shuo Xiao}
\affil{Guizhou Provincial Key Laboratory of Radio Astronomy and Data Processing, 
Guizhou Normal University, Guiyang 550001, China}
\affil{School of Physics and Electronic Science, Guizhou Normal University, Guiyang 550001, China}

\author{Wen-Jun Tan}
\affiliation{State Key Laboratory of Particle Astrophysics, Institute of High Energy Physics, Chinese Academy of Sciences, Beijing 100049, China}
\affiliation{University of Chinese Academy of Sciences, Chinese Academy of Sciences, Beijing 100049, China}

\author{Chao Zheng}
\affiliation{State Key Laboratory of Particle Astrophysics, Institute of High Energy Physics, Chinese Academy of Sciences, Beijing 100049, China}
\affiliation{University of Chinese Academy of Sciences, Chinese Academy of Sciences, Beijing 100049, China}

\author{Jia-Cong Liu}
\affiliation{State Key Laboratory of Particle Astrophysics, Institute of High Energy Physics, Chinese Academy of Sciences, Beijing 100049, China}
\affiliation{University of Chinese Academy of Sciences, Chinese Academy of Sciences, Beijing 100049, China}

\author{Sheng-Lun Xie}
\affil{State Key Laboratory of Particle Astrophysics, Institute of High Energy Physics, Chinese Academy of Sciences, Beijing 100049, China}
\affil{Institute of Astrophysics, Central China Normal University, Wuhan 430079, China}

\author{Peng Zhang}
\affil{State Key Laboratory of Particle Astrophysics, Institute of High Energy Physics, Chinese Academy of Sciences, Beijing 100049, China}
\affil{College of Electronic and Information Engineering, Tongji University, Shanghai 201804, China}

\author{Wen-Long Zhang}
\affil{State Key Laboratory of Particle Astrophysics, Institute of High Energy Physics, Chinese Academy of Sciences, Beijing 100049, China}
\affil{School of Physics and Physical Engineering, Qufu Normal University, Qufu, Shandong 273165, China}

\author{Yue Wang}
\affiliation{State Key Laboratory of Particle Astrophysics, Institute of High Energy Physics, Chinese Academy of Sciences, Beijing 100049, China}
\affiliation{University of Chinese Academy of Sciences, Chinese Academy of Sciences, Beijing 100049, China}

\author{Zheng-Hang Yu}
\affiliation{State Key Laboratory of Particle Astrophysics, Institute of High Energy Physics, Chinese Academy of Sciences, Beijing 100049, China}
\affiliation{University of Chinese Academy of Sciences, Chinese Academy of Sciences, Beijing 100049, China}

\author{Yang-Zhao Ren}
\affiliation{State Key Laboratory of Particle Astrophysics, Institute of High Energy Physics, Chinese Academy of Sciences, Beijing 100049, China}
\affiliation{School of Physical Science and Technology, Southwest Jiaotong University, Chengdu 611756, Sichuan, China}

\author{Ping Wang}
\affil{State Key Laboratory of Particle Astrophysics, Institute of High Energy Physics, Chinese Academy of Sciences, Beijing 100049, China}

\author{Yue Huang}
\affil{State Key Laboratory of Particle Astrophysics, Institute of High Energy Physics, Chinese Academy of Sciences, Beijing 100049, China}

\author{Xiao-Bo Li}
\affil{State Key Laboratory of Particle Astrophysics, Institute of High Energy Physics, Chinese Academy of Sciences, Beijing 100049, China}

\author{Xiao-Yun Zhao}
\affil{State Key Laboratory of Particle Astrophysics, Institute of High Energy Physics, Chinese Academy of Sciences, Beijing 100049, China}

\author{Shi-Jie Zheng}
\affil{State Key Laboratory of Particle Astrophysics, Institute of High Energy Physics, Chinese Academy of Sciences, Beijing 100049, China}

\author{Zhen Zhang}
\affil{State Key Laboratory of Particle Astrophysics, Institute of High Energy Physics, Chinese Academy of Sciences, Beijing 100049, China}

\author{Shu-Xu Yi}
\affil{State Key Laboratory of Particle Astrophysics, Institute of High Energy Physics, Chinese Academy of Sciences, Beijing 100049, China}

\author{Li-Ming Song}
\affil{State Key Laboratory of Particle Astrophysics, Institute of High Energy Physics, Chinese Academy of Sciences, Beijing 100049, China}

\author{Shuang-Nan Zhang}
\affil{State Key Laboratory of Particle Astrophysics, Institute of High Energy Physics, Chinese Academy of Sciences, Beijing 100049, China}

\begin{abstract}
As a soft X-ray imager with unprecedentedly large field of view, EP/WXT has detected many (fast) X-ray transients, whose nature is very intriguing. Whether there is gamma-ray counterpart for the X-ray transient provides important implications for its origin. 
Some of them have been reported to be associated with GRB, however, a systematic study on the gamma-ray emission of these X-ray transients is lacking. In this work, we implemented a comprehensive targeted search for gamma-ray counterparts to 63 X-ray transients reported by EP/WXT during its first year of operation, using the dedicated multiple-instrument search pipeline, ETJASMIN, with GECAM-B, GECAM-C, Fermi/GBM, and \textit{Insight}-HXMT data. We find that 14 out of 63 (22\%) EP/WXT X-ray transients have gamma-ray counterparts. For other transients, ETJASMIN pipeline provided upper limit of gamma-ray emission,
which is more stringent than that given by individual instrument. Moreover, we investigated the properties of the X-ray transients and their gamma-ray counterparts, and explored the relation between the x-ray transient and gamma-ray counterpart.
\end{abstract}


\section{Introduction} \label{sec:intro}
The universe contains a wealth of X-ray transient sources, including (but not limited to) 
X-ray Flashes (XRFs) \citep[e.g.][]{John2001grba.conf...16H,2005NCimC..28..339S,2007A&A...465L..13G},
X-ray counterparts of Gamma-Ray Bursts (GRBs)—characterized as X-ray rich, low-luminosity, and off-axis sources \citep[e.g.][]{Yin2024ApJ...975L..27Y,Liu2024arXiv240416425L},
Soft Gamma-ray Repeaters (SGRs) \citep[e.g.][]{Rea11Magnetar,Olausen14MagnetarCatalog,Mereghetti15Magnetar,Turolla15Magnetar,Kaspi17Magnetar,Esposito21Magnetar},
tidal disruption events (TDEs) \citep[e.g.][]{1988Natur.333..523R,2021ARA&A..59...21G},
X-ray binaries (XRBs) \citep[e.g.][]{Shakura1973A&A....24..337S,2006csxs.book.....L,Remillard2006ARA&A..44...49R,2012MmSAI..83..230C,2016AdSpR..58..761R,2019NewAR..8601546K},
supernova shock breakouts \citep[e.g.][]{2017hsn..book..967W,2020ApJ...896...39A},
and flares from late-type stars \citep[e.g.][]{Glennie2015MNRAS.450.3765G,2015A&A...581A..28P}, 
etc.
X-ray transient sources exhibit a broad range of radiation durations, from sub-seconds to several years. The study of transient phenomena of X-ray sources provides insight into high-energy astrophysical processes, e.g., black hole accretion, binary star evolution and stellar evolution. This information is invaluable for understanding the mysteries of the universe and provides a substantial contribution to the multi-wavelength and multi-messenger era.

As one major source of these transients, GRBs are generally classified into two groups based on their durations $\rm T_{90}$ \citep{1993ApJ...413L.101K}: short GRBs with $\rm T_{90} \textless 2$ s and long GRBs with $\rm T_{90} \textgreater 2$ s. A more physically motivated scheme was proposed by \citet{2007ApJ...655L..25Z,2009ApJ...703.1696Z}. In this framework, Type I GRBs are associated with compact object mergers (e.g., \citealt{Short_GRB}); while Type II GRBs originate from the core collapse of massive stars (e.g., \citealt{Long_GRB}). Generally, short GRBs correspond to Type I and long GRBs to Type II, with some exceptions such as short long GRB \citep{2021NatAs...5..911Z} and long short GRB \citep{2022Natur.612..232Y,2024ApJ...970....6X} as well as the subclass Type IL GRB \citep{2025ApJ...979...73W}.

Several GRB missions are currently in operation, such as $Fermi$/GBM (8-40 MeV, \cite{meegan2009fermi}), $Swift$/BAT (15-150 keV, \cite{suzuki2003hard}), Konus-WIND (20 keV-15 MeV, \cite{1995SSRv...71..265A,2022ApJS..262...32L}), \HXMT /HE (80 keV- 3 MeV, \cite{2020SCPMA..6349503L}), and GECAM (10 keV-6 MeV, \cite{li2022technology,ZHANG2023168586,2024SCPMA..6711013F}). Espcially, the GECAM mission has been enlarged to contain a series of satellites and instruments in the low Earth orbit (LEO) and distant retrograde orbit (DRO). 

Despite of abundant observations on GRB prompt emission extending from tens of keV to GeV, prompt emission observation in the soft X-ray band (keV) have been relatively scarce over the past few decades \footnote{In this work, soft X-ray refers to photons with energies less than $\sim$ 5 keV, while gamma-ray refers to photons with energies greater than $\sim$ 5 keV. The latter is also often referred to as hard X-ray and/or soft gamma-ray.}.
The observations in soft X-ray band have achieved major progress since the launch of the Einstein Probe (EP, \cite{2022cosp...44.1966Y,2024SPIE13093E..1CY}), which is a space mission dedicated to studying the time-domain X-ray astronomy. It is equipped with a Wide-field X-ray Telescope (WXT, \cite{EP/WXT}) utilizing lobster-eye micro-pore optics technology for soft X-ray sky monitoring, as well as a narrow-field Follow-up X-ray Telescope (FXT, \cite{EP/FXT}) for pointed observations. 

As a pathfinder of EP/WXT, the Lobster Eye Imager for Astronomy (LEIA, \cite{LEIA}) instrument detected the prompt emission of GRB 230307A in the 0.5-4 keV energy range. The soft X-ray emission observed by LEIA reveals distinct properties from the gamma-ray emission (from tens of keV to several MeV) deteted by GECAM, which suggests the presence of a magnetar as the central engine of GRB 230307A \citep{GRB230307A_sunhui}. This highlights the importance of joint observations of GRB prompt emission in both the gamma-ray band and soft X-ray band. 

Since its launch, EP/WXT has captured many (fast) X-ray transient events and promptly issued relevant observational information through astronomical alert systems such as ATel\footnote{\url{https://www.astronomerstelegram.org/}} and GCN\footnote{\url{https://gcn.nasa.gov/circulars}}, and has initiated numerous follow-up multi-wavelength observations for various transient events.

Various X-transients have been detected since the launch of EP, such as fast blue optical transients (LFBOTs) \citep{2025ApJ...982L..47V}, jetted tidal disruption events \citep{2025ApJ...979L..30O}, 
GRB \citep{Yin2024ApJ...975L..27Y,Liu2024arXiv240416425L,2025arXiv250304306J}, and X-ray binaries \citep{2025ApJ...980L..36M}, etc.
Despite of many studies on individual X-ray transient, the nature of all these X-ray transients is very intriguing. Whether there is gamma-ray (high-energy) counterpart for soft X-ray transients is crucial to understanding their nature. Currently, only a limited number of gamma-ray counterparts have been reported for the X-ray transients detected by EP/WXT, including the first GRB detected by EP, GRB 240219A \citep{Yin2024ApJ...975L..27Y}, 
the high-redshift GRB 240315a \citep{Liu2024arXiv240416425L}, and a low-energy, extremely soft 
GRB\citep{2025arXiv250304306J}.
However, these studies 
have only utilized individual gamma-ray monitor, such as Fermi/GBM, GECAM, Konus-Wind, or Swift/BAT. Consequently, the detection sensitivity and/or coverage of these gamma-ray searches are constrained.

The Energetic Transients joint analysis system for Multi-INstrument (ETJASMIN) adopts the coherent search algorithm and combine observational data from multiple instruments, which can help us make more effective use of the detection information from multiple satellites \citep{2025ApJS..277....9C}. Relevant studies have shown that ETJASMIN can produce higher search significance compared with a single instrument, and also has better classification and identification capabilities, which is particularly suitable for detecting short and/or faint bursts.

In this work, we perform a systematic targeted search for gamma-ray counterparts of the soft X-ray transients reported by EP in the year of 2024, with the ETJASMIN pipeline using GECAM-B, GECAM-C, Fermi/GBM, and \HXMT instruments. The detailed introduction of these instruments is presented in Section \ref{sec:instru}. 
Detailed observation information, search strategy and data analysis in Section \ref{sec:basic}. The search results and analysis of gamma-ray counterparts are presented in Section \ref{sec:results}, followed by a summary in Section \ref{sec:summary}.

\section{Search Pipeline and Instruments} \label{sec:pipelineinstru}

\subsection{ETJASMIN} \label{sec:ETJASMIN}

ETJASMIN is a dedicated pipeline designed to process and incorporate data from multiple instruments 
\citep{2022MNRAS.514.2397X,2025ApJS..277....9C}. 
It provides a comprehensive analytical framework for studying energetic transients, including gamma-ray bursts, soft gamma-ray repeaters, solar flares, and other high-energy events. With its comprehensive design, the pipeline supports diverse analyses critical to the study of gamma-ray transients, including burst search, verification, classification, positional, spectral and temporal analyses of bursts.

For the gamma-ray counterpart search for EP/WXT transients in this work, a coherent likelihood ratio-based method \citep{2021MNRAS.508.3910C,2023MNRAS.518.2005C,2025ApJS..277....9C,2025SCPMA..6839511C} is employed in ETJASMIN to jointly using data from GECAM-B, GECAM-C and Fermi/GBM, which have similar energy response. We note that, the joint method of ETJASMIN achieves higher significance of burst detection. In addition, ETJASMIN pipeline can effectively verify and classify sub-threshold triggers by jointly analyzing data from multiple missions, which is challenging to achieve using individual instrument alone (see \cite{2025ApJS..277....9C} for more details).

In addition, ETJASMIN employs time-delay localization using the Li-CCF method, effectively utilizing the temporal information in high-resolution data to highlight the signal of interest and deliver more precise results. Again, GECAM-B, GECAM-C and Fermi/GBM have similar energy responses, facilitating spectral fitting, spectral lag analysis, and minimum variability time-scale studies. Under the ETJASMIN framework, joint analysis of these instruments have been demonstrated to be able to yield more accurate and reliable results (see \cite{2022MNRAS.514.2397X} for more details).

\subsection{Instruments} \label{sec:instru}
In the section, we briefly introduce the instruments we used in this work.

Gravitational wave high-energy Electromagnetic Counterpart All-sky Monitor (GECAM) is a series of constellations designed for the detection of GRBs (e.g. \cite{HXMT-GECAM:GRB221009A,Emission_Lines,Afterglow_zhengchao,GRB230307A_sunhui,mini_jet_yishuxu}) , soft gamma-ray repeaters (SGRs, e.g. \cite{Minimum_Variation_Timescales_xiao}) , solar flares (SFs, e.g. \cite{quasi-periodic_SF_zhaohs}), X-ray Binary bursts (XRBs, e.g. \cite{XRBs}), terrestrial gamma-ray flashes and terrestrial electron beams (TGFs and TEBS, e.g. \cite{TGF_TEB_zhaoyi}) and other high-energy gamma transient sources, in the energy range of approximately 10 keV to 6 MeV. 

The GECAM constellation currently comprises four on-orbit members.
GECAM-A and GECAM-B are the first two microsatellites \citep{li2022technology} launched in December 10, 2020. Following the design of the detectors on the GECAM-A and GECAM-B, the High Energy Burst Researcher (i.e. GECAM-C)\citep{ZHANG2023168586} was launched onboard the SATech-01 satellite in July 27, 2022. All these three satellites were launched to LEO, while the GTM instrument (i.e. GECAM-D) was launched to DRO in March 13, 2024 on board the DRO-A satellite \citep{2024SCPMA..6711013F,2024ExA....57...26W}. Owing to various reasons, GECAM-A and GECAM-D have not been in regular observation phase, we only use GECAM-B and GECAM-C data in the current study.

GECAM payloads are equipped with Gamma-ray Detectors (GRDs, \citep{an2022design}), which are capable of detecting both hard X-ray and gamma-ray energy band. GECAM-A, GECAM-B, and GECAM-C satellites are equipped with Charged Particle Detectors (CPDs, \citep{xu2022design}) designed for the detection of charged particles. The simultaneous use of two type detectors can help us determine whether an outburst event is a X/$\rm \gamma$-ray burst or a charged particle event.

\HXMT (Hard X-ray Modulation Telescope) is a satellite designed for X/$\rm \gamma$-ray astronomy, with a broad energy range from 1 keV to 3000 keV \citep{2020SCPMA..6349502Z}. Among its three main instruments, the High Energy Telescope (HE, \cite{2020SCPMA..6349503L}) comprises 18 NaI(Tl)/CsI(Na) phoswich detectors as the main detector. 
The CsI detectors of HE serve dual purposes: as an anti-coincidence system to suppress background signals for NaI and as a gamma-ray monitor to extend the detection capabilities to higher energies. Each CsI crystal has a diameter of 190 mm, and all 18 CsIs provide a total geometric area of approximately 5100 $\rm cm^2$. The detectable energy range for the CsI detectors is about 80-800 keV in normal mode and 200 keV-3 MeV (deposited energy) in low-gain mode \citep{2020SCPMA..6349503L}. The large effective area and wide energy coverage make the CsI detectors essential for studying high-energy transient phenomena \citep{HXMT-GECAM:GRB221009A,2022ApJS..259...46S,2022ApJS..260...24C,2024ApJ...970....6X,2025SCPMA..6851013Y}. Moreover, there is ACD detectors equipped for HE to monitor particle events and label veto signal for NaI/CsI events \citep{2020SCPMA..6349503L}.

The Fermi Gamma-ray Burst Monitor (Fermi/GBM) is an instrument aboard the Fermi Gamma-ray Space Telescope. Equipped with 12 NaI(Tl) and 2 BGO detectors, Fermi/GBM has an effective detection energy range spanning from around 8 keV to 40 MeV \citep{meegan2009fermi}. The data from Fermi/GBM are promptly public in its website \footnote{\url{https://heasarc.gsfc.nasa.gov/FTP/fermi/data/gbm/}}, facilitating convenient access for this targeted search.

\section{Gamma-ray counterpart search and analysis} \label{sec:basic}

\subsection{X-ray transient sample} \label{sec:x-ray transients}
During the first year of operation (i.e. year of 2024), a total of 63 X-ray transients have been reported by EP/WXT. This is the sample we studied in this work. Detailed information of these transients is summarized in Table \ref{tab:EP_basic}.
The corresponding trigger alert message numbers are documented in the second column. In addition, the table includes the triggered source name, trigger time, location, peak flux and derived average unabsorbed flux. All fluxes are calculated in the energy range 0.5-4 keV.
The final column presents the consolidated results of ETJASMIN search with multiple gamma-ray instruments, including GECAM, Fermi/GBM and \textit{Insight}-HXMT. 

\subsection{Search Method} \label{sec:search method}

A methodology of coherent search is implemented in the ETJASMIN pipeline \citep{2025ApJS..277....9C}. This method compares the likelihood of a burst signal versus background noise for each detector using background-subtracted data and detector response. By summing log-likelihoods across multiple detectors from different spacecraft, it improves the sensitivity to weak signals.
This pipeline operates in two modes: blind search and targeted search. 

In the current work, thanks to the precise localization accuracy (about 3 arcmin) of the EP/WXT, we employ a targeted search strategy to search for the sources listed in Table \ref{tab:EP_basic} using the observational data from GECAM-B, GECAM-C and Fermi/GBM with ETJASMIN pipeline.
Meanwhile, to bridge the coverage gaps where neither of GECAM-B, GECAM-C and Fermi/GBM was able to observe the source, we performed a separate targeted search with \textit{Insight}-HXMT, utilizing the same method as in the search for gamma-ray bursts and gravitational wave electromagnetic counterparts \citep{2021MNRAS.508.3910C} and SGR J1935+2154 \citep{2022ApJS..260...24C,2022ApJS..260...25C}.

In order to cover the entire burst interval of the EP/WXT X-ray transients, the search time window for the targeted search is set to be 100 s before the EP trigger time (shown in Table \ref{tab:EP_basic}), and the time span after the WXT trigger time is set to be the duration of the X-ray transients reported by EP/WXT. This covers a large enough time window for searching gamma-ray counterparts.

Before commencing the search, we first calculate the visibility of the GECAM-B, GECAM-C, Fermi/GBM and \HXMT satellites for EP/WXT X-transient sources throughout the search time window. Visible means that the satellite is in operational, data-enabled, not in the South Atlantic Anomaly (SAA), and its field of view is not blocked by the Earth (because these instruments can basically monitor all-sky unblocked by the Earth). 

\begin{figure*}[http]
\centering
  \centerline{
\includegraphics[width=1.6\columnwidth]{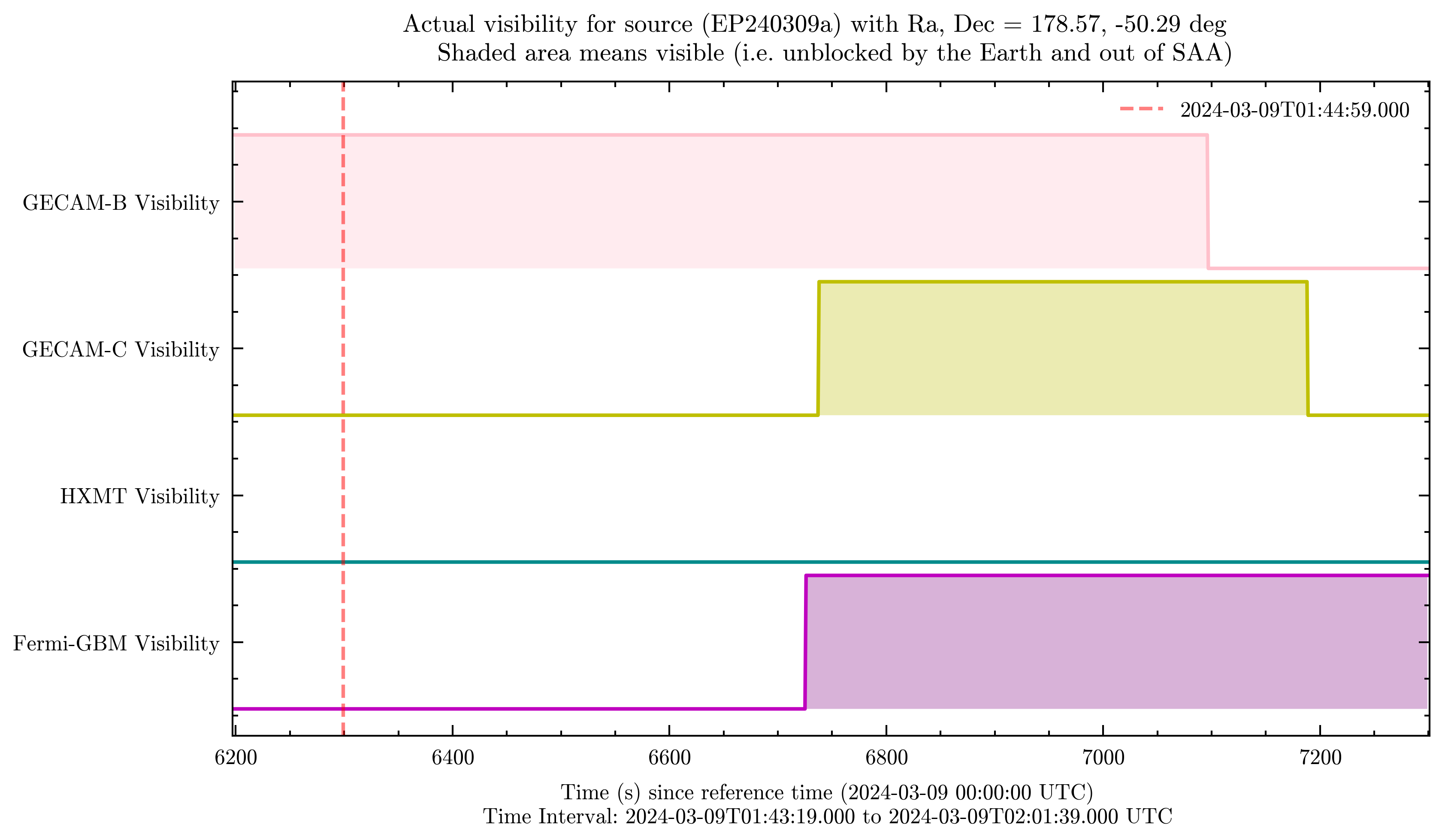}
      }
  \caption{The schematic of visibility of each satellite, which refers to the time period when data are available (instrument is turned on and the satellite is not in SAA area) and the source is in the field of view of the instrument. It should be noted that GECAM-C also excludes high-latitude regions. The shaded area represents the time interval that is visible to the instrument.} 
  \label{fig:visible_figure}
\end{figure*}

\begin{figure*}[http]
\centering
  \centerline{
      \includegraphics[width=0.95\columnwidth]{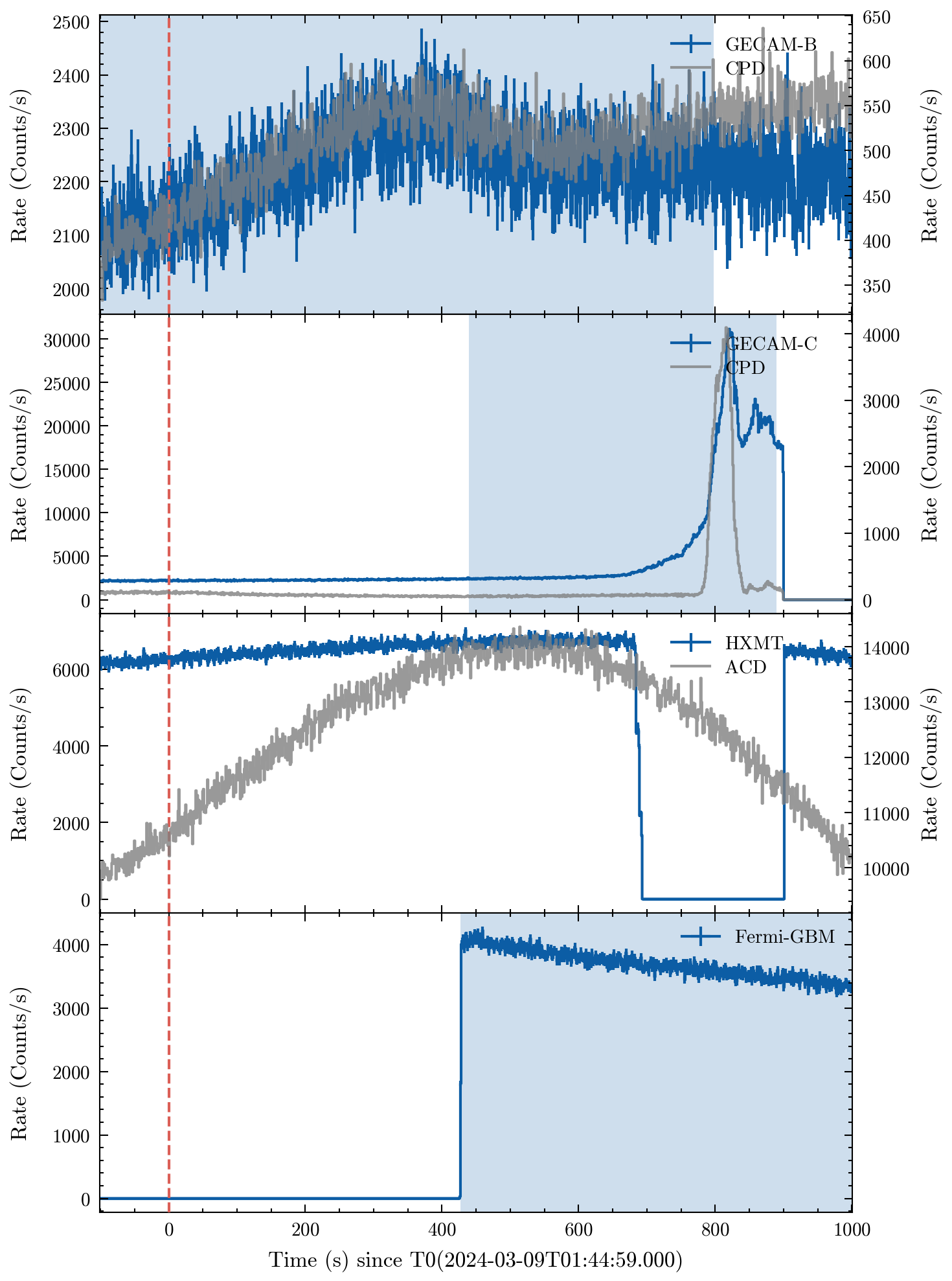}
      }
  \caption{Multiple instrument light curves for a X-ray transient. Time zero (vertical red dashed line) is the start time of EP/WXT X-ray transient reported in GCN or Atel.  The shaded regions correspond to the visible time periods of each satellite as depicted in Figure \ref{fig:visible_figure}. The range of x-axis of this plot is the search time window for gamma-ray counterpart. In addition to the main detectors, light curve of ACD detectors are also plotted for GECAM and HXMT.} 
  \label{fig:fitting}
\end{figure*}

The visibility plot for an example of X-ray transient, EP240309a, is shown in Figure \ref{fig:visible_figure}. The dashed red line represents the trigger time of EP/WXT and the shaded area represents the portion of time that is visible by instruments mentioned above.
It can be seen that satellite visibility varies across time periods, which means that we need to select different satellites for the joint search in different time segments according to the visibility. In addition, as shown in Figure \ref{fig:fitting}, we show the light curves detected by different satellites during the search time window. The shaded areas represent the visible intervals of satellite as in Figure \ref{fig:visible_figure}, and the trigger time of EP/WXT is used as the reference time (i.e. time zero) of the light curves. The range of x-axis of Figure \ref{fig:fitting} is the search time window.

According to the situation of visibility, we split the search time window into different time segments and select the corresponding visible satellites in different time segments for targeted search. The search was implemented with seven search time scales: 0.05, 0.1, 0.2, 0.5, 1, 2 and 4 s. The detailed methodology employed in the search process is elucidated in \cite{2025ApJS..277....9C}. In the event of signal detection, the likelihood ratio and significance are computed. For the non-detection case, the 3-sigma upper limit of the flux is calculated from 10 to 1000 keV with the three spectral templates (i.e. hard, normal, 
and soft Band functions) at three time scales (i.e. 0.1 s, 1 s and 10 s), respectively.
As mentioned above, a separate targeted search with \HXMT is employed to supplement the time intervals during which neither of GECAM-B, GECAM-C and Fermi/GBM is visible to the source.

\subsection{Analysis of Gamma-ray counterpart} \label{sec:data analysis}
Upon detection of burst signals during the search, a 50 ms light curve was employed to calculate the $\rm T_{90}$ and $\rm T_{50}$ parameters for each gamma-ray counterpart candidate. Spectrum analysis is then implemented for the $\rm T_{90}$ time interval. In instances where events were too faint to accurately determine $\rm T_{90}$, a manual determination was made to establish a time period covering the entire signal for spectral analysis.

Spectral analysis for each burst is executed for four spectral models, namely the blackbody (bb, eq. \ref{equ:bb_Model}), power-law (pl, eq. \ref{equ:pl_Model}), cutoff power-law (cpl, eq. \ref{equ:Cutoffpl_Model}), and Band function (eq. \ref{equ:band_Model}) \citep{band1993batse}. The best model was identified based on the Akaike information criterion (AIC) and the Bayesian information criterion (BIC) \citep{2007MNRAS.377L..74L}. The analysis tool \texttt{\textit{Elisa}}\footnote{\url{https://astro-elisa.readthedocs.io/en/latest/notebooks/quick-start.html}}, \footnote{\url{https://github.com/wcxve/elisa}} based on \texttt{\textit{python}} is used in this work.

The blackbody model is defined as
\begin{equation}  
N_{\rm bb}(E)=K\frac{C(E^2)}{\exp(E/kT)-1},
\label{equ:bb_Model}
\end{equation}
where $C=1.0344 \times 10^{-3}$ $\rm photons \cdot cm^{-2} \cdot s^{-1} \cdot keV^{-3}$, $K=R_{km}^2/D_{10}^2$ is the normalization amplitude constant, $R_{km}^2$ is the source radius in km and $D_{10}^2$ is the luminosity distance to the source in units of 10 kpc, $kT$ is the temperature in units of keV.

The power-law model is
\begin{equation}  
N_{\rm pl}(E)=K(\frac{E}{E_{0}})^{\alpha},
\label{equ:pl_Model}
\end{equation}
where $E_{0}=1$~keV, $K$ is the normalization amplitude constant ($\rm photons \cdot cm^{-2} \cdot s^{-1} \cdot keV^{-1}$) at 1 keV, $\alpha$ is the dimensionless photon index of power law.

The power-law with a high-energy cutoff is
\begin{equation}   
N_{\rm cpl}(E)=K(\frac{E}{E_{0}})^{\alpha} \exp(-\frac{E}{E_{\rm c}}),
\label{equ:Cutoffpl_Model}
\end{equation}
where $E_{0}=1$~keV, $K$ is the normalization amplitude constant ($\rm photons \cdot cm^{-2} \cdot s^{-1} \cdot keV^{-1}$) at 1 keV, $\alpha$ is the dimensionless power law photon index, and $E_{\rm c}$ is the characteristic energy of exponential roll-off in keV.

The Band function is 
\begin{footnotesize}
\begin{equation}
    N_{\rm band}(E)= \begin{cases} 
    K \bigg (\frac{E}{E_{\rm 0}}\bigg )^{\alpha} \exp\bigg(-\frac{E}{E_{\rm c}}\bigg ), (\alpha -\beta)E_{\rm c} \geq E,  \\
    K \bigg [\frac{(\alpha-\beta) E_{\rm c}} {E_{\rm 0}}\bigg ]^{\alpha-\beta} \exp(\beta-\alpha)\bigg (\frac{E}{E_{\rm 0}}\bigg )^{\beta}, (\alpha -\beta)E_{\rm c} \leq E, \\
    E_{\rm 0} = 100 \enspace {\rm keV},
    \end{cases},
    \label{equ:band_Model}
\end{equation}
\end{footnotesize}
where $K$ is the normalization amplitude constant ($\rm photons \cdot cm^{-2} \cdot s^{-1} \cdot keV^{-1}$), $\alpha$ is the dimensionless low-energy power-law index, and $\beta$ is the dimensionless high-energy power-law index, $E_{\rm c}$ is the characteristic energy in keV, $E_{\rm 0}$ is the pivot energy in keV and usually the value is taken as 100 keV.

\subsection{Analysis Results} \label{sec:results}
A detailed targeted search for gamma-ray counterpart for the EP/WXT X-ray transients was carried out using the search methodology described above. 
Search results are summarized in column 10 of Table \ref{tab:EP_basic}. A total of 14 gamma-ray counterpart candidates were found, 
for which detailed search information is documented in Table \ref{tab:search_v1}, including trigger time, trigger time scale, likelihood ratio (LR), significance (SNR), and triggering instruments. 

Light curves of the corresponding gamma-ray counterparts are shown in Figures \ref{fig:search_v1} and \ref{fig:search_v2}, with the blue shaded areas denoting the visible times for each satellite. The red dashed line is the trigger time of the EP, and the purple dashed line is the trigger time of the targeted search. 
We note that, the count rate for CPD of GECAM and ACD of \textit{Insight}-HXMT are shown in grey lines in the light curve plots, 
which can help us identify whether the signal is gamma-ray photon event or charged particle event. 
Additionally, calculations of the corresponding $\rm T_{90}$ and $\rm T_{50}$ values have been performed according to the light curve. 
Time-integrated spectral fitting analysis in the duration of $\rm T_{90}$ period was performed, and the time ranges and the results of the spectral analysis are recorded in Table \ref{tab:search_v2}. The results of the best model fit are also presented in the table, along with the calculated flux from 10 to 1000 keV.

For those X-ray transients that gamma-ray counterparts were not found in the targeted search, we computed the corresponding upper flux limits with ETJASMIN, as shown in Table \ref{tab:search_v0}. The second column shows the time during which the upper limit was calculated, and the column for satellites refers to the satellites used in the upper limit calculation. The upper limit is calculated using three spectral profiles across three time scales. The parameters of the Band spectra for the three spectral profiles (hard, normal, and soft) employed in the calculation of the upper flux limit are as follows:
For the hard Band, the parameters are $E_{\rm peak}$ = 1000 keV, $\alpha$ = 0, $\beta$ = -1.5. 
For the normal Band, the parameters are $E_{\rm peak}$ = 230 keV, $\alpha$ = -1, $\beta$ = -2.3. 
For the soft Band, the parameters are $E_{\rm peak}$ = 70 keV, $\alpha$ = -1.9, $\beta$ = -3.7. 
Note that $E_{\rm peak} = (2+\alpha)* E_{\rm c}$ is defined as the peak characteristic energy in the $\nu F_{\nu}$. 
For the calculation energy range of upper limit flux, it is 10-1000 keV for GECAM and Fermi/GBM. However, for \HXMT operating in the normal mode, since the detectable energy range is relatively narrow, the calculation energy range is 120-600 keV, because the response matrix in this band can describe the data well \citep{2020JHEAp..27....1L,HE_calibration_updated}.

\section{Discussions} \label{sec:Discussion}

\subsection{Properties of X-ray Transients} \label{sec:X-ray Transients}
\begin{figure*}[http]
\centering
  \centerline{
      \includegraphics[width=2.2\columnwidth]{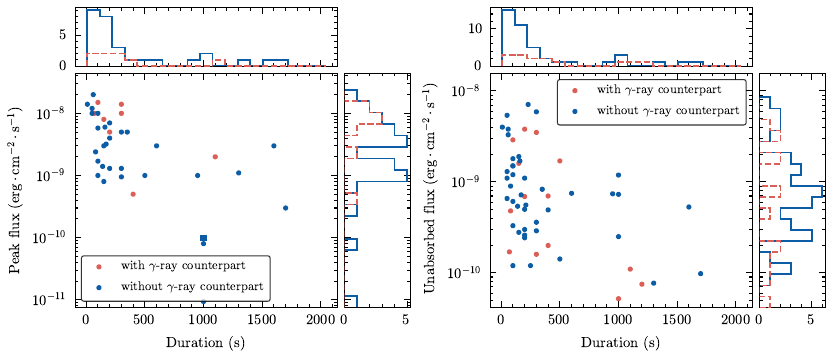}
      }
  \caption{The left and right figures respectively show the flux diagrams of peak flux and derived average flux in the 0.4-4 keV range reported by EP/WXT. The duration is reported by EP/WXT. Among them, the red data points represent the X-ray transients for which gamma-ray counterparts have been found, while the blue points indicate the sources without counterparts.} 
  \label{fig:flux_contrast}
\end{figure*}

We plot the peak flux and the average unabsorbed flux in the 0.5-4 keV range versus duration of X-ray transients measured by EP/WXT in Figure \ref{fig:flux_contrast}. All the data points 
of EP/WXT X-ray transients 
in the figure (also shown in Table \ref{tab:EP_basic})
 are from the EP/WXT observation reports published in the General Coordinates Network (GCN) or the Astronomer's Telegram (ATEL).
The blue data points represent the x-ray transient sources without gamma-ray counterpart, whereas the red data points are the sources with gamma-ray counterpart found in this work.
One can see that the duration of most X-ray transient sources is several hundred seconds. We note that, compared to all x-ray transients, the X-ray transients with gamma-ray counterpart are widely distributed 
in terms of the duration and flux in soft x-ray band. However, the X-ray transients with gamma-rays tend to have higher peak flux in soft X-ray band (Left panel in Figure \ref{fig:flux_contrast}), although their unabsorbed x-ray flux (averaged for the full burst) is evenly distributed without a clear tendency. 

\subsection{Properties of Gamma-ray Counterparts} \label{sec:Gamma-ray Counterparts}

As mentioned before, we calculated $\rm T_{50}$ and $\rm T_{90}$ for these X-ray transients with gamma-ray counterparts. Meanwhile, we generated spectra according to the time range of $\rm T_{90}$ and fitted them with four models described in Section \ref{sec:data analysis}. The best-fitting model was selected based on the Bayesian Information Criterion (BIC, \cite{2007MNRAS.377L..74L}), and the corresponding results were recorded in Tables \ref{tab:search_v1} and \ref{tab:search_v2}. 

\begin{figure*}[http]
\centering
  \centerline{
      \includegraphics[width=2\columnwidth]{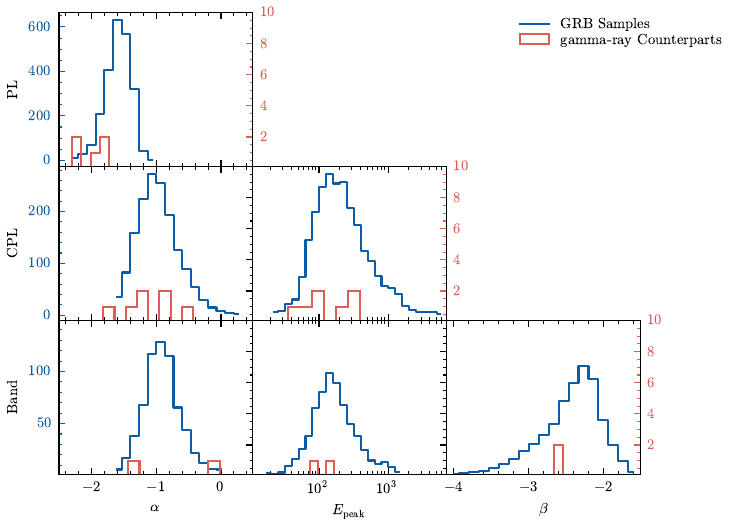}
      }
  \caption{Distribution of spectral parameters of sources with counterparts (red line) in GRB samples. GRB sample data (blue line) are from Fermi/GBM Gamma-Ray Burst Spectral Catalog \citep{2021ApJ...913...60P}. The Comp model is another functional form of the cpl, with $E_{\rm peak}$ instead of $E_{\rm c}$.} 
  \label{fig:GRB_contrast}
\end{figure*}

In addition, we selected the ten-year Gamma-Ray Burst Spectral Catalog of Fermi/GBM \citep{2021ApJ...913...60P} as the comparison sample for gamma-ray counterparts, and the results are shown in Figure \ref{fig:GRB_contrast}. 
The blue lines represent the fitting parameters of the power-law, cutoff power-law, and Band models of normal GRB samples from Fermi/GBM GRB catalog \citep{2021ApJ...913...60P}, respectively. The red lines are for gamma-ray counterparts of EP/WXT X-ray transient found in this work. For each counterpart, only one set of parameters corresponding to the best spectral model is shown.

\begin{figure*}[http]
\centering
  \centerline{
      \includegraphics[width=1.8\columnwidth]{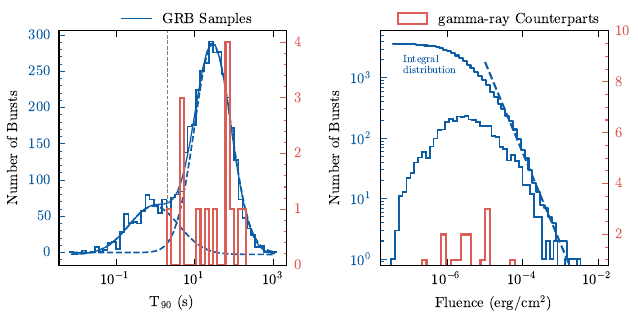}}
\caption{Properties comparison of gamma-ray counterparts with gamma-ray samples. GRB sample data (blue line) are from Fermi/GBM Gamma-Ray Burst Spectral Catalog \citep{2020ApJ...893...46V,2021ApJ...913...60P}}
\label{fig:gammaray_GRBSamples}
\end{figure*}

Based on the results, one can find that the energy spectra of most counterparts can be well-fitted by a cutoff power-law (cpl) model. Regarding the spectral parameters of the power-law (pl) fitting, the spectral index is close to -2, and is located on the left side of the GRB sample. This indicates that the spectrum of the gamma-ray counterpart is relatively soft in the GRB population. For the low-energy spectral index and $E_{\rm peak} = (2+\alpha)*E_{\rm c}$ distribution obtained from cpl and band model fitting results, they are relatively discrete. Regarding the $\beta$ parameter of the Band model, there are only two data points, making it impossible to perform statistical analysis. We note that, the above finding is preliminary, and further statistical analysis requires a larger sample size.

The duration and fluence distributions of the gamma-ray counterparts are also compared with the GRB sample in Figure \ref{fig:gammaray_GRBSamples}, with Fermi/GBM catalog data \citep{2020ApJ...893...46V,2021ApJ...913...60P} shown as reference (blue lines, see also Figure \ref{fig:GRB_contrast}).
We point out that the $\rm T_{90}$ duration distribution shows that all gamma-ray counterparts exceed 2 s, indicating no typical short-GRB counterparts among the X-ray transients detected in 2024. 
Moreover, the durations exhibit a relatively uniform distribution between 2-1000 s, without displaying any distinctive clustering pattern.
The fluence distribution is presented through both differential and integral distribution of the GRB samples. 
The blue dashed line provides visual guidance, with a power law with a slope of -3/2 \citep{2020ApJ...893...46V}.
From this figure, one can see that the fluence distribution of gamma-ray counterparts in red line exhibits remarkable similarity to the overall GRB sample distribution.

\subsection{Relation between X-ray and Gamma-ray emission}\label{sec:xray_garay}
Having compared X-ray transients with and without gamma-ray counterparts, and the gamma-ray counterpart distributions with GRB samples, we now analyze their flux, fluence, and duration characteristics of X-ray transients associated with gamma-ray counterparts.

\begin{figure}[http]
\gridline{\fig{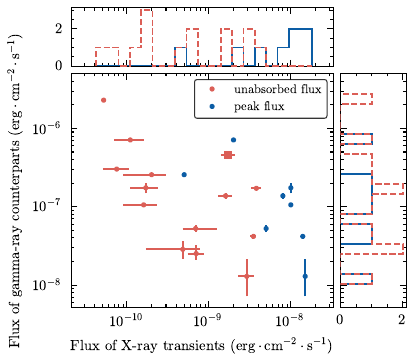}{0.42\textwidth}{(a) flux}}
\gridline{\fig{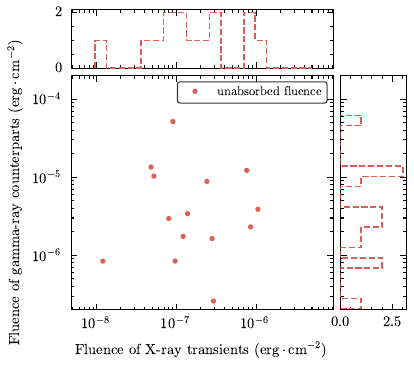}{0.42\textwidth}{(b) fluence}
}
\gridline{\fig{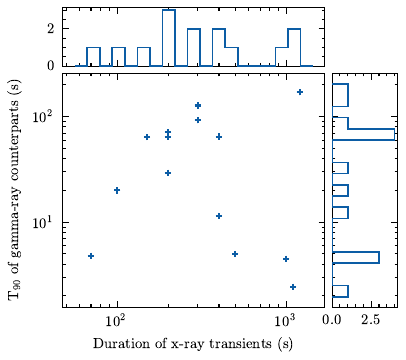}{0.42\textwidth}{(c) duration}
}
\caption{Properties comparison of gamma-ray counterparts with X-ray transient.}
\label{fig:flux_xray_garay}
\end{figure}

As shown in panel a of Figure \ref{fig:flux_xray_garay}, an inverse correlation is evident between the derived average unabsorbed 0.5-4 keV flux of X-ray transients and the flux of their gamma-ray counterparts within the $\rm T_{90}$ time period. 
The Pearson correlation coefficient was -0.592, and the corresponding $p$-value was determined to be 0.026. This result indicates the significant of 2.23 $\sigma$.
However, this correlation is not seen in the fluence plot (panel b of Figure \ref{fig:flux_xray_garay}).

As can be seen from the plot of the durations of the X-ray transients against their gamma-ray counterparts (denoted by $\rm T_{90}$), there is no clear correlation exhibited.
It is noteworthy that the durations of X-ray transients are typically much longer than those of their gamma-ray counterparts.

\subsection{Gamma-ray Instrument Visibility}\label{sec:Visibility statistics}

\begin{figure*}[http]
\centering
  \centerline{
      \includegraphics[width=2.1\columnwidth]{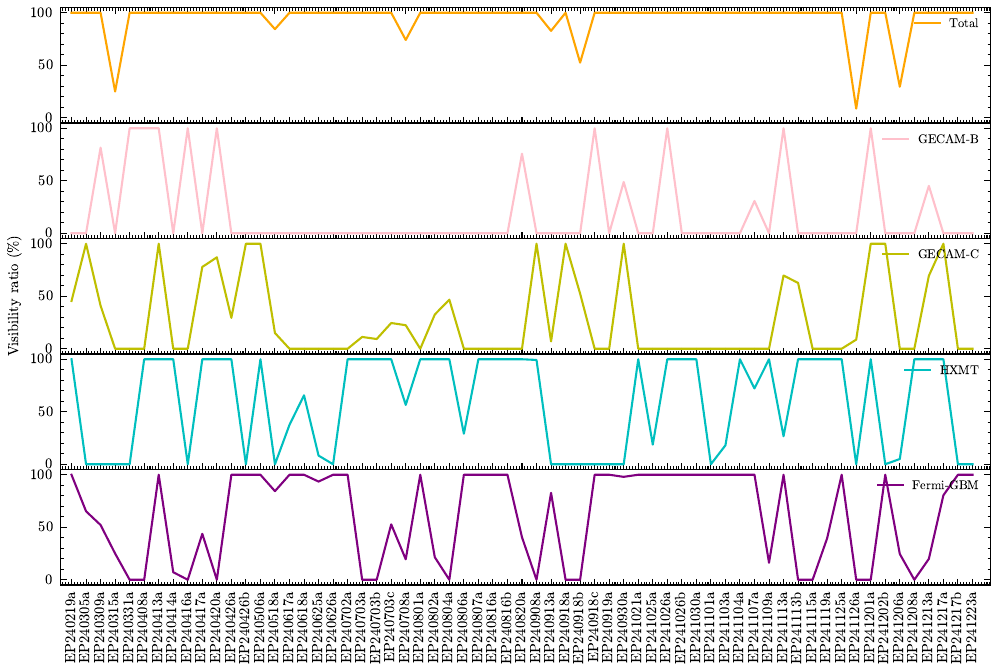}
      }
  \caption{Ratio of the visibility of each satellite for each X-ray transient over the full search time range.} 
  \label{fig:Ratio_each_burst_v0}
\end{figure*}

As an important aspect of multiple instrument search, the visibility of each instrument determine many factors of the search result. 
As shown in Figure \ref{fig:visible_figure}, the visibility of each satellite for each X-ray transient varies over different time periods. A single satellite can monitor the X-ray transient only during certain time intervals. Therefore, for each X-ray transient, we can estimate the visibility ratio of each monitor, as well as the ratio after combining multiple monitors. The specific results are presented in Figure \ref{fig:Ratio_each_burst_v0}. It can be observed that achieving full time coverage of all X-ray transients is very difficult for a single satellite. However, by combining multiple devices, full monitoring coverage capabilities can be achieved for most transients.

\begin{figure*}[http]
\centering
  \centerline{
      \includegraphics[width=2.2\columnwidth]{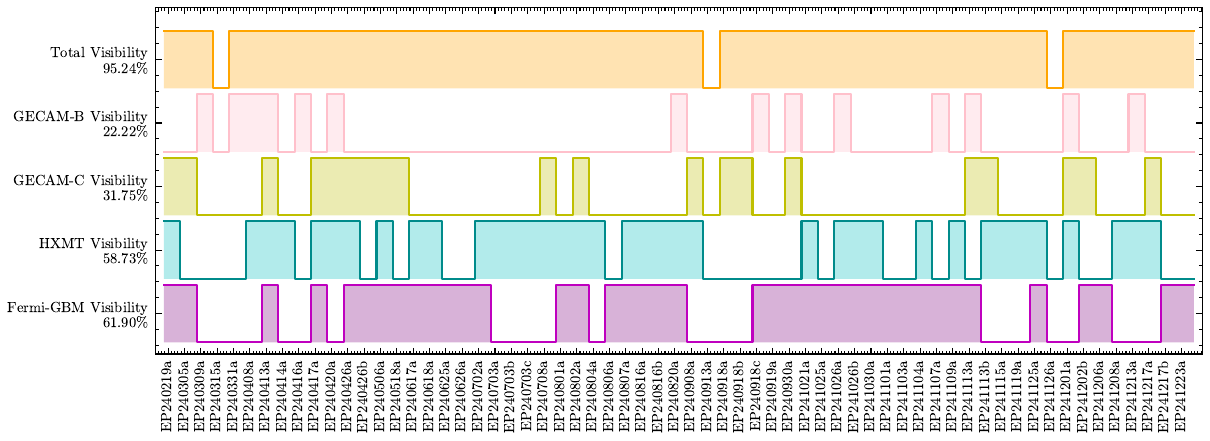}
      }
  \caption{Visibility of each satellite to the X-ray transient at trigger time. Percentage represents how much of all X-ray transient sources are visible. Shaded areas represent visible.} 
  \label{fig:Ratio_each_satellite_v0}
\end{figure*}

We calculated the visibility and coverage rate of each X-ray transient, and the corresponding results are displayed in Figure \ref{fig:Ratio_each_satellite_v0} and Figure \ref{fig:Ratio_each_burst_v0}.
In Figure \ref{fig:Ratio_each_satellite_v0}, 
the shaded areas represent the time when the instruments are visible. One can find that, considering Fermi/GBM alone, only 61.90\% of X-ray transients can be monitored during the burst time. However, after adding GECAM-B, GECAM-C, and \textit{Insight}-HXMT, 95.24\% of the sources can be monitored by at least one of these telescopes. 
This significantly improves the monitoring coverage rate and demonstrates the 
advantages of our research.

\section{Summary} \label{sec:summary}
Searching for gamma-ray counterparts of X-ray transients detected by EP/WXT is of great significance. Some search results have been released on GCN and ATEL by many individual gamma-ray monitors, such as Fermi/GBM, GECAM, Konus-Wind, INTEGRAL/SPI-ACS, Swift/BAT, etc. However, the field of view, data collection time, detection sensitivity of each of these instruments are limited. 

Therefore, in this paper we aim to combine multiple instruments to achieve a comprehensive search for gamma-ray counterparts of EP/WXT X-ray transients using the Energetic Transients joint analysis system for Multi-INstrument (ETJASMIN).
We first collected all X-ray transient sources detected by EP/WXT in the year of 2024, then conducted a targeted search for gamma-ray counterparts using the ETJASMIN pipeline.

Thanks to the data of GECAM-B, GECAM-C, Fermi/GBM, and \textit{Insight}-HXMT satellites, we can achieve observational coverage for 95\% of the X-ray transients at the moment of their triggering, allowing for a more comprehensive targeted search with the ETJASMIN pipeline. Additionally, due to the relay observations from multiple satellites, we can achieve the full search coverage capabilities of 100 seconds before the trigger and duration after the burst for most X-ray transient samples, which can be considered the most complete search. 

After the systematic search, we found gamma-ray counterpart candidate for 14 out of 63 (22\%) X-ray transients. The search results demonstrate that the instantaneous soft X-ray almost always precedes its gamma-ray counterpart and exhibits a significantly longer duration. For those x-ray transients without gamma-ray counterparts found, the most stringent upper limit of flux in the 10-1000 keV range was calculated using data from multiple instruments by ETJASMIN.

While most of these counterparts have previously been reported on GCN, here we systematically performed calculations of $\rm T_{90}$ for these counterparts and conducted joint spectral analyses during this time period with data of multiple instruments. 

Interestingly, we notice that all gamma-ray counterparts of these x-ray transients last longer than 2 seconds, a traditional threshold for short and long GRB. Until this work, the unique merger-origin GRB detected by EP/WXT or its pathfinder LEIA is the peculiar long-short GRB 230307A \citep{GRB230307A_sunhui,mini_jet_yishuxu}. 

We further investigate the properties of the soft x-ray transients and their gamma-ray counterparts.
We find that the X-ray transients with gamma-ray counterpart tend to have higher peak flux in soft x-ray band, while their duration and averaged flux have no preference in the distribution compared to the sample of all X-ray transients. 
The spectrum of the gamma-ray counterparts is relatively soft, but $\rm T_{90}$ and fluence don't show a peculiar distribution. 

For these 14 X-ray transients with gamma-ray counterparts, we also compared the flux, fluence, and duration between X-ray and gamma-ray band, and we find that an inverse correlation is evident between the 0.5-4 keV flux of X-ray transients and the fluxes of their gamma-ray counterparts.

Lastly, we note that the current sample size of the current work is rather small, however, these findings may have important implications for the physics of the x-ray transients as well as the gamma-ray counterparts.

\section*{Acknowledgments}
This work is supported by the National Key R\&D Program of China (Grant No. 2022YFF0711404, 
2021YFA0718500), 
the National Natural Science Foundation of China 
(Grant No. 12273042, 
12303045,
12494572
),
the Natural Science Foundation of Hebei Province (No. A2023205020) 
and the Strategic Priority Research Program of the Chinese Academy of Sciences (Grant No. XDB0550300, 
XDA30050000). 
The GECAM (Huairou-1) mission is supported by the Strategic Priority Research Program on Space Science of the Chinese Academy of Sciences (XDA15360000). 
The \HXMT missioni is funded by the China National Space Administration (CNSA) and the Chinese Academy of Sciences (CAS).
We appreciate the development and operation teams of GECAM, SATech-01, \HXMT and EP.
We acknowledge the public data and software from Fermi/GBM and Intergral/SPI-ACS, the alert data from EP and the software from \texttt{\textit{Elisa}}, \texttt{\textit{GECAMTools}}, and \texttt{\textit{GBM Data Tools}}.

\newpage
\bibliography{sample631.bib}{}
\bibliographystyle{aasjournal}

\newpage
\begin{longrotatetable}
\begin{deluxetable*}{cccccccccc}
\centering
\tablecaption{Basic observations of X-ray transient detected by Einstein Probe\label{tab:EP_basic}}
\renewcommand{\arraystretch}{0.6}
\tablewidth{700pt}
\tabletypesize{\scriptsize}
\tablehead{
\colhead{Source Name} & \colhead{Alert Number} & 
\colhead{Trigger Time} & \colhead{Duration} & 
\colhead{Ra} & \colhead{Dec} & 
\colhead{Err} & \colhead{Peak Flux} & 
\colhead{Average Unabsorbed flux} & \colhead{Gamma-Ray Counterpart}\\ 
\colhead{} & \colhead{} & \colhead{(UTC)} & \colhead{(s)} & 
\colhead{($^{\circ}$)} & \colhead{($^{\circ}$)} & \colhead{($^{\prime}$)} &
\colhead{($\rm erg \cdot cm^{-2} \cdot s^{-1}$)} & \colhead{($\rm erg \cdot cm^{-2} \cdot s^{-1}$)} & \colhead{}
} 
\renewcommand{\arraystretch}{1}
\colnumbers
\startdata
EP240219a & ATel 16463 & 2024-02-19T06:21:42.000 & $\sim$200 & 80.0 & 25.5 & 3.0 & $5 \times 10^{-9} $ & $6.9 ^{+5.6}_{-2.1} \times 10^{-10}$ & yes \\
EP240305a & ATel 16509 & 2024-03-05T14:15:31.000 & $\sim$350 & 122.9 & -54.7 & 3.0 & $5 \times 10^{-9} $ & $8.3 ^{+2.0}_{-1.4} \times 10^{-10}$ & no \\
EP240309a & ATel 16546 & 2024-03-09T01:44:59.000 & 1000$^{\dagger}$ & 178.6 & -50.3 & 2.1 & $9.40 \times 10^{-12} $ & - & no \\
EP240315a & GCN 35931 & 2024-03-15T20:10:44.000 & $\sim$1600 & 141.6 & -9.5 & 3.0 & $3 \times 10^{-9} $ & $5.3 ^{+1.0}_{-0.7} \times 10^{-10}$ & no \\
EP240331a & ATel 16564 & 2024-03-31T22:07:17.000 & $\sim$100 & 169.4 & -20.0 & 20.0 & - & $1.8 ^{+0.3}_{-0.3} \times 10^{-9}$ & no \\
EP240408a & GCN 36053 & 2024-04-08T17:56:30.000 & $\sim$10 & 158.8 & -35.7 & 3.0 & $1.4 \times 10^{-8} $ & $4.0 ^{+1.3}_{-1.3} \times 10^{-9}$ & no \\
EP240413a & GCN 36086 & 2024-04-13T14:39:37.000 & $\sim$200 & 228.8 & -18.8 & 20.0 & $7 \times 10^{-9} $ & $1.1 ^{+0.1}_{-0.1} \times 10^{-9}$ & no \\
EP240414a & GCN 36091 & 2024-04-14T09:50:12.000 & $\sim$150 & 191.5 & -9.7 & 3.0 & $3 \times 10^{-9} $ & - & no \\
EP240416a & GCN 36138 & 2024-04-16T02:42:13.000 & $\textgreater$200 & 203.2 & -13.6 & 3.0 & $1.3 \times 10^{-9} $ & $5.0 ^{+2.6}_{-1.6} \times 10^{-10}$ & no \\
EP240417a & GCN 36161 & 2024-04-17T15:12:33.000 & $\textgreater$1700 & 177.4 & -15.4 & 3.0 & $3 \times 10^{-10} $ & $9.8 ^{+0.8}_{-0.8} \times 10^{-11}$ & no \\
EP240420a & GCN 36194 & 2024-04-20T12:04:28.000 & $\textgreater$100 & 228.7 & 14.8 & 3.0 & $1 \times 10^{-8} $ & - & no \\
 EP240426a & GCN 36313 & 2024-04-26T04:23:21.000 & $\sim$534 & 121.9 & -29.5 & 0.2 & - & - & no \\
EP240426b & GCN 36330 & 2024-04-26T14:19:06.000 & $\sim$300 & 173.8 & -40.7 & 3.0 & $9.5 \times 10^{-10} $ & $3.6 ^{+0.6}_{-0.6} \times 10^{-10}$ & no \\
EP240506a & GCN 36405 & 2024-05-06T05:01:39.000 & $\sim$50 & 214.0 & -16.7 & 3.0 & $1 \times 10^{-8} $ & $1.1 ^{+0.3}_{-0.2} \times 10^{-9}$ & no \\
EP240518a & GCN 36512 & 2024-05-18T13:01:50.000 & $\textgreater$1000 & 217.0 & -49.6 & 3.0 & $8 \times 10^{-11} $ & - & no \\
EP240617a & GCN 36691 & 2024-06-17T12:19:13.000 & $\textgreater$300 & 285.0 & -22.6 & 3.0 & $1.4 \times 10^{-8} $ & $3.5 ^{+0.3}_{-0.3} \times 10^{-9}$ & yes \\
EP240618a & GCN 36690 & 2024-06-18T05:43:43.000 & $\sim$100 & 281.6 & 23.8 & 3.0 & $1.5 \times 10^{-8} $ & $2.9 ^{+0.7}_{-0.6} \times 10^{-9}$ & yes \\
EP240625a & GCN 36757 & 2024-06-25T01:48:23.000 & $\sim$300 & 310.8 & -16.0 & 2.0 & $1.3 \times 10^{-9} $ & $2.9 ^{+0.7}_{-0.5} \times 10^{-10}$ & no \\
EP240626a & GCN 36766 & 2024-06-26T06:28:28.000 & $\sim$160 & 263.0 & -13.1 & 2.0 & $6 \times 10^{-9} $ & $1.70 ^{+0.73}_{-0.50} \times 10^{-9}$ & no \\
EP240702a & GCN 36801 & 2024-07-02T00:50:05.000 & $\sim$50 & 328.2 & -39.0 & 3.0 & $1.2 \times 10^{-8} $ & $5.4 ^{+1.5}_{-1.2} \times 10^{-9}$ & no \\
EP240703a & GCN 36807 & 2024-07-03T00:38:40.000 & $\sim$300 & 273.8 & -9.7 & 3.0 & $5 \times 10^{-9} $ & $5.9 ^{+8.6}_{-2.6} \times 10^{-9}$ & no \\
EP240703b & GCN 36810 & 2024-07-03T05:24:26.000 & $\sim$600 & 279.5 & -57.4 & 3.0 & $3 \times 10^{-9} $ & $7.5 ^{+1.3}_{-1.8} \times 10^{-10}$ & no \\
EP240703c & GCN 36818 & 2024-07-03T18:15:00.000 & $\textgreater$1000 & 289.3 & -30.3 & 3.0 & - & $2.5 ^{+0.8}_{-0.8} \times 10^{-10}$ & no \\
EP240708a & GCN 36838 & 2024-07-08T23:28:23.000 & $\sim$1300 & 346.0 & -22.8 & 3.0 & $1.1 \times 10^{-9} $ & $7.7 ^{+3.4}_{-2.6} \times 10^{-11}$ & no \\
EP240801a & GCN 36997 & 2024-08-01T09:06:20.700 & $\textgreater$200 & 345.1 & 32.6 & 2.4 & - & $4.8 ^{+3.1}_{-3.1} \times 10^{-10}$ & yes \\
EP240802a & GCN 37019 & 2024-08-02T10:32:52.000 & $\textgreater$500 & 287.8 & -2.4 & 1.9 & - & $1.7 ^{+0.4}_{-0.3} \times 10^{-9}$ & yes \\
EP240804a & GCN 37034 & 2024-08-04T21:36:53.000 & $\textgreater$100 & 337.6 & -39.1 & 4.0 & - & $6.1 ^{+2.6}_{-1.6} \times 10^{-10}$ & no \\
EP240806a & GCN 37063 & 2024-08-06T04:47:53.900 & $\sim$150 & 11.5 & 5.1 & 2.1 & - & $1.9 ^{+1.8}_{-0.6} \times 10^{-9}$ & no \\
EP240807a & GCN 37088 & 2024-08-07T05:42:48.000 & $\sim$70 & 301.0 & -68.8 & 2.5 & $1 \times 10^{-8} $ & $1.7 ^{+0.7}_{-0.6} \times 10^{-10}$ & yes \\
EP240816a & GCN 37188 & 2024-08-16T03:28:42.000 & $\textgreater$200 & 292.9 & -54.4 & 2.7 & - & $3.0 ^{+0.5}_{-0.5} \times 10^{-10}$ & no \\
EP240816b & GCN 37185 & 2024-08-16T01:44:27.000 & $\textgreater$50 & 16.0 & 15.4 & 3.3 & - & $1.3 ^{+0.7}_{-0.5} \times 10^{-9}$ & no \\
EP240820a & GCN 37214 & 2024-08-20T00:54:47.000 & $\sim$250 & 16.2 & -34.7 & 2.1 & - & $1.2 ^{+0.7}_{-0.5} \times 10^{-10}$ & no \\
EP240908a & GCN 37432 & 2024-09-08T17:28:27.000 & $\sim$950 & 14.0 & 8.1 & 2.7 & $1 \times 10^{-9} $ & $7.4 ^{+2.4}_{-1.9} \times 10^{-10}$ & no \\
EP240913a & GCN 37492 & 2024-09-13T11:39:33.000 & $\sim$1100 & 16.7 & 16.8 & 2.5 & $2 \times 10^{-9} $ & $1.1 ^{+0.5}_{-0.4} \times 10^{-10}$ & yes \\
EP240918a & GCN 37554 & 2024-09-18T11:21:52.000 & $\sim$170 & 289.4 & 46.1 & 0.3 & $3.2 \times 10^{-9} $ & $7.2 ^{+2.8}_{-1.9} \times 10^{-10}$ & no \\
EP240918b & GCN 37555 & 2024-09-18T15:40:00.000 & $\sim$200 & 258.7 & 66.7 & 2.9 & - & $2.6 ^{+4.1}_{-1.2} \times 10^{-10}$ & no \\
EP240918c & GCN 37555 & 2024-09-18T18:06:47.000 & $\sim$100 & 281.3 & -13.2 & 2.3 & $5.8 \times 10^{-9} $ & $1.5 ^{+0.6}_{-0.2} \times 10^{-9}$ & no \\
EP240919a & GCN 37561 & 2024-09-19T14:47:40.000 & $\textgreater$400 & 334.3 & -9.7 & 0.3 & - & $7.0 ^{+1.8}_{-1.5} \times 10^{-10}$ & yes \\
EP240930a & GCN 37648 & 2024-09-30T17:17:50.000 & $\sim$200 & 319.9 & 41.3 & 2.1 & - & $3.8 ^{+0.5}_{-0.5} \times 10^{-9}$ & yes \\
EP241021a & GCN 37834 & 2024-10-21T05:07:56.000 & $\sim$100 & 28.9 & 6.0 & 2.4 & $1 \times 10^{-9} $ & $3.3 ^{+4.8}_{-1.6} \times 10^{-10}$ & no \\
EP241025a & GCN 37864 & 2024-10-25T01:35:29.000 & $\sim$300 & 333.7 & 83.6 & 0.2 & $1 \times 10^{-8} $ & $1.6 ^{+0.7}_{-0.7} \times 10^{-10}$ & yes \\
EP241026a & GCN 37909 & 2024-10-26T22:41:28.000 & $\sim$150 & 293.4 & 58.0 & 0.2 & $8 \times 10^{-9} $ & $1.6 ^{+0.3}_{-0.3} \times 10^{-9}$ & yes \\
EP241026b & GCN 37902 & 2024-10-26T18:14:30.000 & $\sim$100 & 56.4 & 41.0 & 2.9 & $1.7 \times 10^{-9} $ & $1.2 ^{+0.4}_{-0.3} \times 10^{-10}$ & no \\
EP241030a & GCN 37997 & 2024-10-30T06:33:18.000 & $\sim$1200 & 343.0 & 80.4 & 2.4 & - & $7.5 ^{+3.0}_{-2.4} \times 10^{-11}$ & yes \\
EP241101a & GCN 38039 & 2024-11-01T23:52:49.000 & $\sim$100 & 37.8 & 22.7 & 2.8 & - & $1.2 ^{+0.5}_{-0.4} \times 10^{-9}$ & no \\
EP241103a & GCN 38051 & 2024-11-03T01:23:37.000 & $\sim$60 & 27.8 & 19.0 & 0.2 & - & $3.8 ^{+1.1}_{-0.9} \times 10^{-9}$ & no \\
EP241104a & GCN 38081 & 2024-11-04T18:34:15.425 & $\sim$400 & 32.6 & 31.6 & 2.7 & $5.0 \times 10^{-10} $ & $2.0 ^{+1.0}_{-1.1} \times 10^{-10}$ & yes \\
EP241107a* & GCN 38112 & 2024-11-07T14:10:23.000 & 1000$^{\dagger}$ & 35.0 & 3.3 & 0.2 & $1 \times 10^{-10} $ & - & no \\
EP241109a* & GCN 38140 & 2024-11-09T06:01:55.000 & 1000$^{\dagger}$ & 18.4 & 0.0 & 0.2 & - & - & no \\
EP241113a & GCN 38211 & 2024-11-13T19:09:19.000 & $\textgreater$210 & 132.0 & 52.4 & 0.2 & - & $5.57 ^{+1.26}_{-0.76} \times 10^{-10}$ & no \\
EP241113b & GCN 38214 & 2024-11-13T11:22:50.000 & $\sim$100 & 110.2 & 46.8 & 2.5 & - & $1.5 ^{+0.4}_{-0.4} \times 10^{-9}$ & no \\
EP241115a & GCN 38239 & 2024-11-15T05:47:20.000 & $\sim$500 & 19.4 & -18.0 & 2.6 & $1 \times 10^{-9} $ & $1.42 ^{+0.6}_{-0.4} \times 10^{-10}$ & no \\
EP241119a & GCN 38281 & 2024-11-19T17:53:20.000 & $\sim$200 & 84.1 & 3.8 & 2.3 & $4 \times 10^{-9} $ & $2.43 ^{+0.67}_{-0.52} \times 10^{-10}$ & no \\
EP241125a & GCN 38318 & 2024-11-25T00:06:06.000 & $\textgreater$150 & 48.6 & 37.7 & 2.6 & $8 \times 10^{-10} $ & $2.79 ^{+1.11}_{-0.86} \times 10^{-10}$ & no \\
EP241126a & GCN 38335 & 2024-11-26T19:39:41.000 & $\textgreater$60 & 3.7 & 11.7 & 2.4 & $2 \times 10^{-8} $ & $3.3 ^{+0.9}_{-0.7} \times 10^{-9}$ & no \\
EP241201a & GCN 38415 & 2024-12-01T20:59:16.000 & $\sim$230 & 282.6 & 66.1 & 2.3 & - & $7.1 ^{+32.7}_{-4.9} \times 10^{-9}$ & no \\
EP241202b & GCN 38426 & 2024-12-02T15:12:55.000 & $\textgreater$140 & 45.3 & 2.4 & 2.6 & $1.4 \times 10^{-9} $ & $5.4 ^{+2.0}_{-1.6} \times 10^{-10}$ & no \\
EP241206a & GCN 38457 & 2024-12-06T16:34:47.000 & $\sim$400 & 34.7 & 38.9 & 3.8 & - & $4.92 ^{+1.19}_{-1.24} \times 10^{-10}$ & no \\
EP241208a & GCN 38477 & 2024-12-08T16:36:13.000 & 1000$^{\dagger}$ & 127.8 & 49.1 & 4.0 & - & $6.57 ^{+4.73}_{-2.49} \times 10^{-10}$ & no \\
EP241213a & GCN 38554 & 2024-12-13T02:17:15.000 & 1000$^{\dagger}$ & 116.2 & 35.3 & 2.8 & - & $5.2  \times 10^{-11}$ & yes \\
EP241217a & GCN 38586 & 2024-12-17T05:36:03.000 & 1000$^{\dagger}$ & 47.0 & 30.9 & 2.8 & - & $7.3 ^{+2.7}_{-2.7} \times 10^{-10}$ & no \\
EP241217b & GCN 38606 & 2024-12-17T16:58:04.349 & 1000$^{\dagger}$ & 84.2 & -25.3 & 2.8 & - & $1.19 ^{+0.10}_{-0.10} \times 10^{-9}$ & no \\
EP241223a & GCN 38660 & 2024-12-23T07:21:23.000 & $\sim$80 & 74.8 & 7.1 & 3.3 & $2.4 \times 10^{-9} $ & $9.0 ^{+4.0}_{-4.0} \times 10^{-10}$ & no \\
\enddata
\vspace{-0.1cm}
\tablecomments{All fluxes are calculated in the 0.5 to 4 keV energy range. When source Name is followed by *, it means it's a stellar flare event. The duration followed by $^{\dagger}$ indicates no corresponding duration were published, which we define as 1000 s to ensure a sufficient search time window.
}
\end{deluxetable*}
\end{longrotatetable}


\setlength\tabcolsep{4pt}
\begin{deluxetable*}{cccccccccccc}
\centering
\tablecaption{Multi-INstrument (ETJASMIN) Search Upper Flux Limit Results\label{tab:search_v0}}
\tablewidth{0pt}
\tabletypesize{\scriptsize}
\renewcommand{\arraystretch}{0.6}
\tablewidth{100pt}
\tabletypesize{\scriptsize}
\tablehead{
\colhead{Source Name} & 
 \colhead{Upper Limit Time} & \colhead{Satellites} &  \multicolumn{9}{c}{Upper Limit}\\ 
\cmidrule(r){4-7}
\cmidrule(r){8-12}
& \colhead{(UTC)} & 
\colhead{} & \multicolumn{4}{c}{($10^{-7}  \cdot \rm erg \cdot cm^{-2} \cdot s^{-1}$) }&  \multicolumn{5}{c}{($10^{-8}  \cdot \rm erg \cdot cm^{-2} \cdot s^{-1}$)}
} 
\vspace{-1cm}
\renewcommand{\arraystretch}{1}
\colnumbers
\startdata
EP240305a & 2024-03-05T14:15:31.000 & Fermi/GBM,GECAM-C & 3.58 & 1.90 & 1.21 & 1.13 & 6.01 & 3.84 & 3.58 & 1.90 & 1.21 \\
EP240309a & 2024-03-09T01:52:23.000 & GECAM-B & 3.12 & 1.84 & 1.32 & 0.99 & 5.82 & 4.17 & 3.12 & 1.84 & 1.32 \\
EP240315a & 2024-03-15T20:32:19.000 & Fermi/GBM & 4.71 & 2.22 & 1.35 & 1.49 & 7.01 & 4.27 & 4.71 & 2.22 & 1.35 \\
EP240331a & 2024-03-31T22:07:17.000 & GECAM-B & 9.46 & 5.50 & 5.07 & 2.99 & 17.40 & 16.04 & 9.46 & 5.50 & 5.07 \\
EP240408a & 2024-04-08T17:55:30.000 & GECAM-B & 8.43 & 4.69 & 4.24 & 2.67 & 14.82 & 13.41 & 8.43 & 4.69 & 4.24 \\
EP240413a & 2024-04-13T14:39:37.000 & Fermi/GBM,GECAM-B,GECAM-C & 3.21 & 1.76 & 1.25 & 1.01 & 5.57 & 3.96 & 3.21 & 1.76 & 1.25 \\
EP240414a & 2024-04-14T09:53:05.000 & \HXMT & 2.85 & 8.90 & 9.37 & 0.90 & 28.14 & 29.63 & 2.85 & 8.90 & 9.37 \\
EP240416a & 2024-04-16T02:42:13.000 & Fermi/GBM,GECAM-B,GECAM-C & 7.82 & 5.09 & 5.06 & 2.47 & 16.11 & 15.99 & 7.82 & 5.09 & 5.06 \\
EP240417a & 2024-04-17T15:12:33.000 & Fermi/GBM,GECAM-C & 3.19 & 1.84 & 1.30 & 1.01 & 5.83 & 4.12 & 3.19 & 1.84 & 1.30 \\
EP240420a & 2024-04-20T12:04:28.000 & GECAM-B,GECAM-C & 14.79 & 12.11 & 12.99 & 4.68 & 38.30 & 41.09 & 14.79 & 12.11 & 12.99 \\
EP240426a & 2024-04-26T04:23:21.000 & Fermi/GBM,GECAM-C & 4.76 & 2.33 & 1.50 & 1.50 & 7.37 & 4.75 & 4.76 & 2.33 & 1.50 \\
EP240426b & 2024-04-26T14:19:06.000 & Fermi/GBM,GECAM-C & 3.22 & 1.84 & 1.25 & 1.02 & 5.82 & 3.95 & 3.22 & 1.84 & 1.25 \\
EP240506a & 2024-05-06T05:01:39.000 & Fermi/GBM,GECAM-C & 3.79 & 2.06 & 1.39 & 1.20 & 6.50 & 4.39 & 3.79 & 2.06 & 1.39 \\
EP240518a & 2024-05-18T13:04:42.000 & Fermi/GBM & 4.34 & 2.25 & 1.41 & 1.37 & 7.12 & 4.46 & 4.34 & 2.25 & 1.41 \\
EP240625a & 2024-06-25T01:48:23.000 & Fermi/GBM & 5.36 & 3.02 & 1.94 & 1.70 & 9.56 & 6.15 & 5.36 & 3.02 & 1.94 \\
EP240626a & 2024-06-26T06:28:28.000 & Fermi/GBM & 4.83 & 2.68 & 1.75 & 1.53 & 8.49 & 5.53 & 4.83 & 2.68 & 1.75 \\
EP240702a & 2024-07-02T00:50:05.000 & Fermi/GBM & 4.09 & 2.20 & 1.31 & 1.29 & 6.97 & 4.14 & 4.09 & 2.20 & 1.31 \\
EP240703a & 2024-07-03T00:38:40.000 & \HXMT & 1.89 & 6.84 & 8.03 & 0.60 & 21.65 & 25.38 & 1.89 & 6.84 & 8.03 \\
EP240703b & 2024-07-03T05:24:26.000 & \HXMT & 1.48 & 1.90 & 1.69 & 0.47 & 6.01 & 5.34 & 1.48 & 1.90 & 1.69 \\
EP240703c & 2024-07-03T18:22:40.000 & Fermi/GBM,GECAM-C & 4.25 & 2.38 & 1.50 & 1.35 & 7.53 & 4.74 & 4.25 & 2.38 & 1.50 \\
EP240708a & 2024-07-08T23:28:23.000 & \HXMT & 3.05 & 5.42 & 5.08 & 0.97 & 17.15 & 16.05 & 3.05 & 5.42 & 5.08 \\
%
%
EP240804a & 2024-08-04T21:36:53.000 & \HXMT & 3.10 & 6.34 & 6.00 & 0.98 & 20.06 & 18.98 & 3.10 & 6.34 & 6.00 \\
EP240806a & 2024-08-06T04:47:53.900 & GECAM-C & 3.70 & 2.03 & 1.36 & 1.17 & 6.43 & 4.30 & 3.70 & 2.03 & 1.36 \\
EP240816a & 2024-08-16T03:28:42.000 & Fermi/GBM & 7.12 & 5.43 & 5.80 & 2.25 & 17.17 & 18.35 & 7.12 & 5.43 & 5.80 \\
EP240816b & 2024-08-16T01:44:27.000 & Fermi/GBM & 4.24 & 2.06 & 1.25 & 1.34 & 6.53 & 3.96 & 4.24 & 2.06 & 1.25 \\
EP240820a & 2024-08-20T00:54:47.000 & \HXMT & 2.14 & 3.04 & 2.74 & 0.68 & 9.61 & 8.68 & 2.14 & 3.04 & 2.74 \\
EP240908a & 2024-09-08T17:28:27.000 & GECAM-C & 4.99 & 2.87 & 2.38 & 1.58 & 9.07 & 7.52 & 4.99 & 2.87 & 2.38 \\
EP240918a & 2024-09-18T11:21:52.000 & GECAM-C & 5.07 & 3.45 & 2.96 & 1.60 & 10.91 & 9.35 & 5.07 & 3.45 & 2.96 \\
EP240918b & 2024-09-18T15:40:00.000 & GECAM-C & 6.17 & 3.09 & 1.89 & 1.95 & 9.79 & 5.97 & 6.17 & 3.09 & 1.89 \\
EP240918c & 2024-09-18T18:06:47.000 & Fermi/GBM,GECAM-B & 3.22 & 1.72 & 1.26 & 1.02 & 5.43 & 3.98 & 3.22 & 1.72 & 1.26 \\
EP241021a & 2024-10-21T05:07:56.000 & Fermi/GBM & 4.00 & 2.11 & 1.35 & 1.26 & 6.67 & 4.28 & 4.00 & 2.11 & 1.35 \\
EP241026b & 2024-10-26T18:14:30.000 & Fermi/GBM & 3.85 & 2.16 & 1.42 & 1.22 & 6.82 & 4.50 & 3.85 & 2.16 & 1.42 \\
EP241101a & 2024-11-01T23:52:49.000 & Fermi/GBM & 3.76 & 1.93 & 1.24 & 1.19 & 6.10 & 3.92 & 3.76 & 1.93 & 1.24 \\
EP241103a & 2024-11-03T01:23:37.000 & Fermi/GBM & 3.69 & 2.05 & 1.34 & 1.17 & 6.48 & 4.24 & 3.69 & 2.05 & 1.34 \\
EP241107a* & 2024-11-07T14:10:23.000 & Fermi/GBM,GECAM-B & 3.94 & 2.04 & 1.28 & 1.25 & 6.46 & 4.05 & 3.94 & 2.04 & 1.28 \\
EP241109a* & 2024-11-09T06:01:55.000 & Fermi/GBM & 4.49 & 2.65 & 1.72 & 1.42 & 8.37 & 5.44 & 4.49 & 2.65 & 1.72 \\
EP241113a & 2024-11-13T19:09:19.000 & Fermi/GBM,GECAM-B,GECAM-C & 3.28 & 1.71 & 1.15 & 1.04 & 5.41 & 3.63 & 3.28 & 1.71 & 1.15 \\
EP241113b & 2024-11-13T11:22:50.000 & \HXMT & 2.35 & 7.64 & 8.22 & 0.74 & 24.16 & 25.99 & 2.35 & 7.64 & 8.22 \\
EP241115a & 2024-11-15T05:47:20.000 & \HXMT & 2.49 & 4.57 & 4.30 & 0.79 & 14.46 & 13.59 & 2.49 & 4.57 & 4.30 \\
EP241119a & 2024-11-19T17:53:20.000 & \HXMT & 2.18 & 3.65 & 3.40 & 0.69 & 11.53 & 10.74 & 2.18 & 3.65 & 3.40 \\
EP241125a & 2024-11-25T00:06:06.000 & Fermi/GBM & 4.39 & 2.21 & 1.34 & 1.39 & 6.97 & 4.23 & 4.39 & 2.21 & 1.34 \\
EP241126a & 2024-11-26T19:39:41.000 & - & - & - & - & - & - & - & - & - & - \\
EP241201a & 2024-12-01T20:59:16.000 & GECAM-B,GECAM-C & 3.72 & 1.93 & 1.57 & 1.18 & 6.09 & 4.97 & 3.72 & 1.93 & 1.57 \\
EP241202b & 2024-12-02T15:12:55.000 & Fermi/GBM,GECAM-C & 3.32 & 1.78 & 1.15 & 1.05 & 5.63 & 3.64 & 3.32 & 1.78 & 1.15 \\
EP241206a & 2024-12-06T16:34:47.000 & Fermi/GBM & 4.84 & 2.75 & 1.70 & 1.53 & 8.70 & 5.37 & 4.84 & 2.75 & 1.70 \\
EP241208a & 2024-12-08T16:36:13.000 & \HXMT & 2.13 & 5.08 & 5.62 & 0.67 & 16.06 & 17.78 & 2.13 & 5.08 & 5.62 \\
EP241217a & 2024-12-17T05:36:03.000 & GECAM-C & 6.02 & 3.60 & 2.45 & 1.90 & 11.38 & 7.76 & 6.02 & 3.60 & 2.45 \\
EP241217b & 2024-12-17T16:58:04.349 & Fermi/GBM & 4.32 & 2.17 & 1.35 & 1.36 & 6.85 & 4.26 & 4.32 & 2.17 & 1.35 \\
EP241223a & 2024-12-23T07:21:23.000 & Fermi/GBM & 4.13 & 2.27 & 1.47 & 1.31 & 7.17 & 4.65 & 4.13 & 2.27 & 1.47 \\
\enddata
\vspace{-0.45cm}
\begin{flushleft}
\tablecomments{The time indicated in column (2) represents the time is at which the upper limit of the flux is calculated. The upper limit flux are calculated in 10 to 1000 keV (GECAM-B,GECCAM-C,Fermi/GBM) and 120 to 600 keV (\HXMT).
And flux are calculated using three time scales and three spectral profiles, 0.1 s: (4)(5)(6), 1 s: (7)(8)(9), 10 s: (10)(11)(12). The three groups under each time scale correspond to hard, normal, and soft band. When source Name is followed by *, it means it's a stellar flare event. } 
\end{flushleft}
\end{deluxetable*}

\setlength\tabcolsep{2pt}
\begin{deluxetable*}{cccccccc}
\centering
\tablecaption{Multi-INstrument (ETJASMIN) Search Results of Gamma-Ray Counterpart Candidates\label{tab:search_v1}}
\tablewidth{0pt}
\tabletypesize{\scriptsize}
\renewcommand{\arraystretch}{0.2}
\tablewidth{10pt}
\tabletypesize{\scriptsize}
\tablehead{
\colhead{Source Name} & \colhead{Trigger Time} & \colhead{Time Scale} & \colhead{LR} & \colhead{Weighted SNR} &  \colhead{Trigger Satellites}  &  \colhead{$\rm T_{90}$} &  \colhead{$\rm T_{50}$}\\ 
& \colhead{(UTC)} & \colhead{(s)} &\colhead{}  & \colhead{}  & \colhead{} & \colhead{(s)} & \colhead{(s)}
} 
\renewcommand{\arraystretch}{0.9}
\colnumbers
\startdata
EP240219a & 2024-02-19T06:21:52.000 & 4.0 & 44.91 & 9.53 & Fermi/GBM & $64.60^{+60.52}_{-24.35}$ & $26.45^{+7.75}_{-7.31}$ \\
EP240617a & 2024-06-17T12:23:00.000 & 4.0 & 102.22 & 14.34 & Fermi/GBM & $92.75^{+25.95}_{-33.63}$ & $27.80^{+4.30}_{-2.95}$ \\
EP240618a & 2024-06-18T05:43:59.000 & 4.0 & 16.12 & 5.81 & Fermi/GBM & - & - \\
EP240801a & 2024-08-01T09:07:47.500 & 2.0 & 13.85 & 5.36 & Fermi/GBM & 
$29.50^{+6.76}_{-9.81}$ & $10.10^{+2.16}_{-1.40}$ \\
EP240802a & 2024-08-02T10:34:05.100 & 2.0 & 85.30 & 12.91 & HXMT & - & - \\
EP240807a & 2024-08-07T05:42:45.250 & 2.0 & 52.46 & 10.28 & HXMT,Fermi/GBM  & - & - \\
EP240913a & 2024-09-13T11:42:35.000 & 2.0 & 1771.91 & 58.43 & Fermi/GBM & $2.45^{+8.98}_{-0.75}$ & $1.00^{+0.65}_{-0.25}$ \\
EP240919a & 2024-09-19T14:49:09.500 & 2.0 & 21.84 & 6.62 & Fermi/GBM & $64.80^{+33.83}_{-16.58}$ & $33.60^{+8.78}_{-11.20}$ \\
EP240930a & 2024-09-30T17:17:58.950 & 4.0 & 1427.35 & 48.8 & Fermi/GBM,GECAM-B,GECAM-C & $70.95^{+7.83}_{-3.25}$ & $39.75^{+3.80}_{-2.58}$ \\
EP241025a & 2024-10-25T01:36:27.000 & 4.0 & 147.01 & 17.13 & Fermi/GBM & $127.65^{+18.50}_{-6.30}$ & $57.70^{+10.50}_{-24.68}$ \\
EP241026a & 2024-10-26T22:41:57.000 & 4.0 & 2121.89 & 64.63 & Fermi/GBM,GECAM-B,HXMT & $64.05^{+23.38}_{-30.13}$ & $10.50^{+1.35}_{-1.75}$ \\
EP241030a & 2024-10-30T05:48:11.600 & 4.0 & 317.75 & 25.22 & Fermi/GBM & $170.20^{+5.90}_{-1.50}$ & $46.05^{+0.45}_{-0.45}$ \\
EP241104a & 2024-11-04T18:30:15.000 & 4.0 & 1571.74 & 40.07 & Fermi/GBM & $11.50^{+5.40}_{-1.60}$ & $4.65^{+0.25}_{-0.30}$ \\
EP241213a & 2024-12-13T02:18:58.250 & 4.0 & 1771.9 & 59.37 & GECAM-B,HXMT & $4.50^{+18.65}_{-1.35}$ & $1.80^{+0.50}_{-0.35}$ \\
\enddata
\vspace{-0.45cm}
\begin{flushleft}
\tablecomments{The time indicated in column (2) represents the trigger time for the detection of a gamma-ray counterpart. EP240802a and EP240930a also detected by INTEGRAL/SPI-ACS.}
\end{flushleft}
\end{deluxetable*}

\setlength\tabcolsep{2pt}
\begin{deluxetable*}{cccccccc}
\tablecaption{Spectrum Fitting Results of Gamma-Ray Counterpart Candidates\label{tab:search_v2}}
\tablewidth{0pt}
\tabletypesize{\scriptsize}
\renewcommand{\arraystretch}{0.8}
\tablewidth{30pt}
\tabletypesize{\scriptsize}
\tablehead{
\colhead{Source Name} & \colhead{Trigger Time} &  \colhead{Time range} & \colhead{$\alpha$} & \colhead{$\beta$} & \colhead{$E_{\rm C}$} & \colhead{Model} & \colhead{ Flux }\\ 
& \colhead{(UTC)} & \colhead{(s)} & \colhead{} & \colhead{} & \colhead{(keV)} & \colhead{}  & \colhead{($\rm erg \cdot cm^{-2} \cdot s^{-1}$)}
} 
\renewcommand{\arraystretch}{0.9}
\colnumbers
\startdata
EP240219a & 2024-02-19T06:21:52.000 & (-3.35,61.25) & $-1.82^{+0.06}_{-0.06}$ & - & - & pl & $5.27^{+0.56}_{-0.52} \times 10^{-8}$ \\
EP240617a & 2024-06-17T12:23:00.000 & (11.4,104.15) & $-1.32^{+0.21}_{-0.18}$ & - & $76.04^{+30.68}_{-20.18}$ & cpl & $4.18^{+0.31}_{-0.25} \times 10^{-8}$ \\
EP240618a & 2024-06-18T05:43:59.000 & (-10,10) & $-2.27^{+0.37}_{-0.59}$ & - & - & pl & $1.29^{+0.89}_{-0.58} \times 10^{-8}$ \\
EP240801a & 2024-08-01T09:07:47.500 & (-11.5,18.0) & $-2.01^{+0.15}_{-0.18}$ & - & - & pl & $2.86^{+0.78}_{-0.68} \times 10^{-8}$ \\
EP240802a & 2024-08-02T10:34:05.000 & (-4,1) & $-2.18^{+0.51}_{-0.92}$ & - & - & pl & $4.61^{+41.43}_{-4.54} \times 10^{-7}$ \\
EP240807a & 2024-08-07T05:42:45.250 & (-0.60,4.20) & $-1.81^{+0.11}_{-0.12}$ & - & - & pl & $1.75^{+0.26}_{-0.26} \times 10^{-7}$ \\
EP240913a & 2024-09-13T11:42:35.000 & (0.85,3.3) & $-0.02^{+0.18}_{-0.23}$ & $-2.63^{+0.17}_{-0.27}$ & $39.41^{+7.18}_{-6.77}$ & Band & $7.13^{+0.44}_{-0.47} \times 10^{-7}$ \\
EP240919a & 2024-09-19T14:49:09.500 & (-5.15,59.65) & $-1.18^{+0.46}_{-0.32}$ & - & $127.13^{+216.26}_{-64.09}$ & cpl & $2.53^{+0.72}_{-0.41} \times 10^{-8}$ \\
EP240930a & 2024-09-30T17:17:58.950 & (0.30,71.25) & $-0.94^{+0.05}_{-0.05}$ & - & $98.31^{+7.34}_{-7.10}$ & cpl & $1.72^{+0.04}_{-0.04} \times 10^{-7}$ \\
EP241025a & 2024-10-25T01:36:27.000 & (-0.80,126.85) & $-0.92^{+0.07}_{-0.07}$ & - & $181.04^{+29.62}_{-24.59}$ & cpl & $1.06^{+0.06}_{-0.05} \times 10^{-7}$ \\
EP241026a & 2024-10-26T22:41:57.000 & (-1.15,62.90) & $-1.14^{+0.09}_{-0.07}$ & - & $324.74^{+118.09}_{-79.74}$ & cpl & $1.38^{+0.13}_{-0.12} \times 10^{-7}$ \\
EP241030a & 2024-10-30T05:48:11.600 & (11.35,181.55) & $-1.34^{+0.02}_{-0.02}$ & $-2.56^{+0.14}_{-0.21}$ & $209.97^{+19.29}_{-18.31}$ & Band & $3.03^{+0.05}_{-0.05} \times 10^{-7}$ \\
EP241104a & 2024-11-04T18:30:15.000 & (-2.95,8.55) & $-0.54^{+0.12}_{-0.10}$ & - & $49.07^{+5.32}_{-4.85}$ & cpl & $2.57^{+0.08}_{-0.07} \times 10^{-7}$ \\
EP241213a & 2024-12-13T02:18:58.250 & (0.40,4.90) & $-1.66^{+0.16}_{-0.13}$ & - & $791.60^{+849.17}_{-332.38}$ & cpl & $2.30^{+0.13}_{-0.15} \times 10^{-6}$ \\
\enddata
\vspace{-0.45cm}
\begin{flushleft}
\tablecomments{The flux in column (8) are calculated in 10 to 1000 keV.}
\end{flushleft}
\end{deluxetable*}

\begin{figure*}
\gridline{\fig{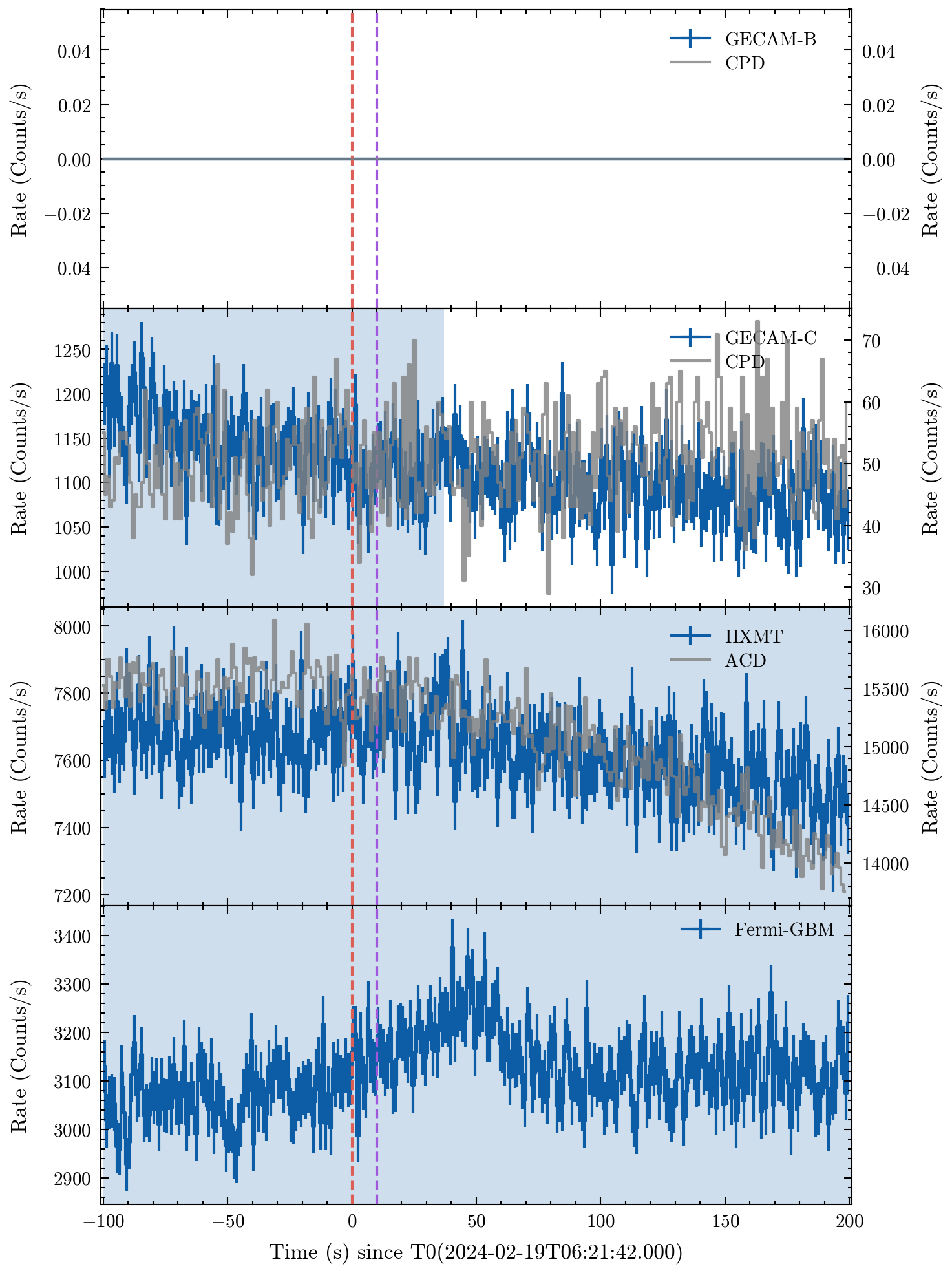}{0.46\textwidth}{(a)}
\fig{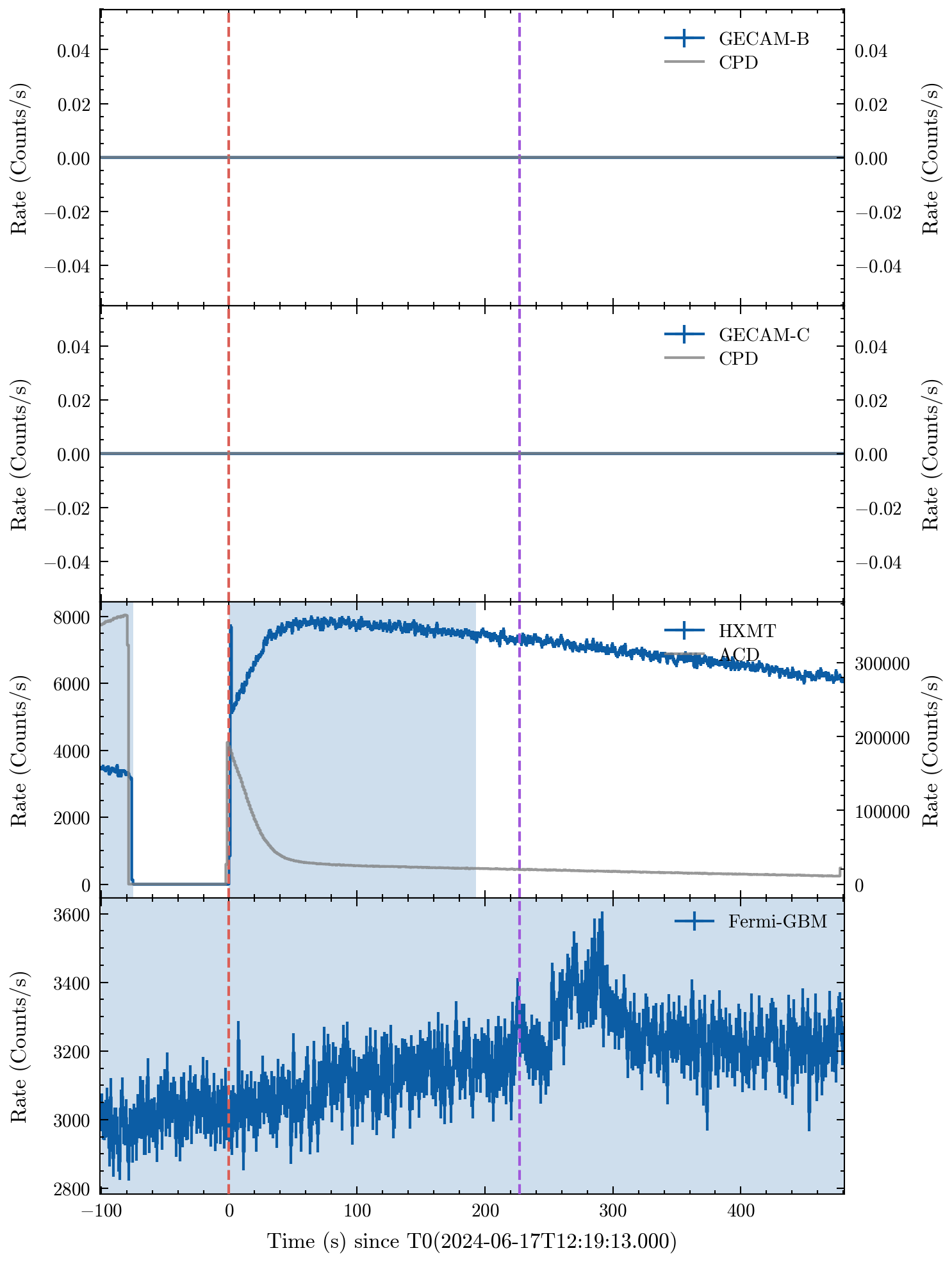}{0.46\textwidth}{(b)}
          }
\renewcommand{\arraystretch}{0.46}
\gridline{
\fig{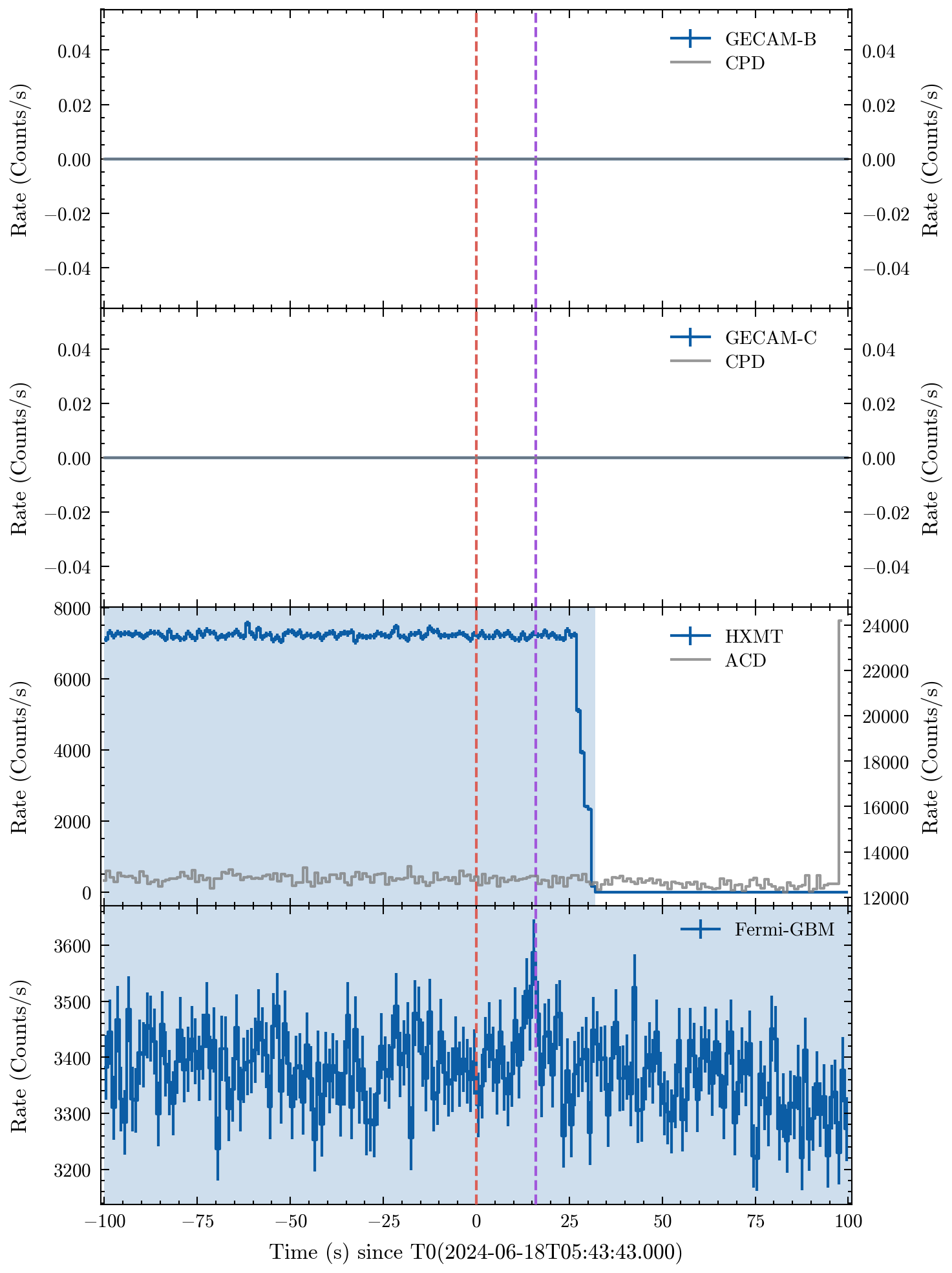}{0.46\textwidth}{(c)}
\fig{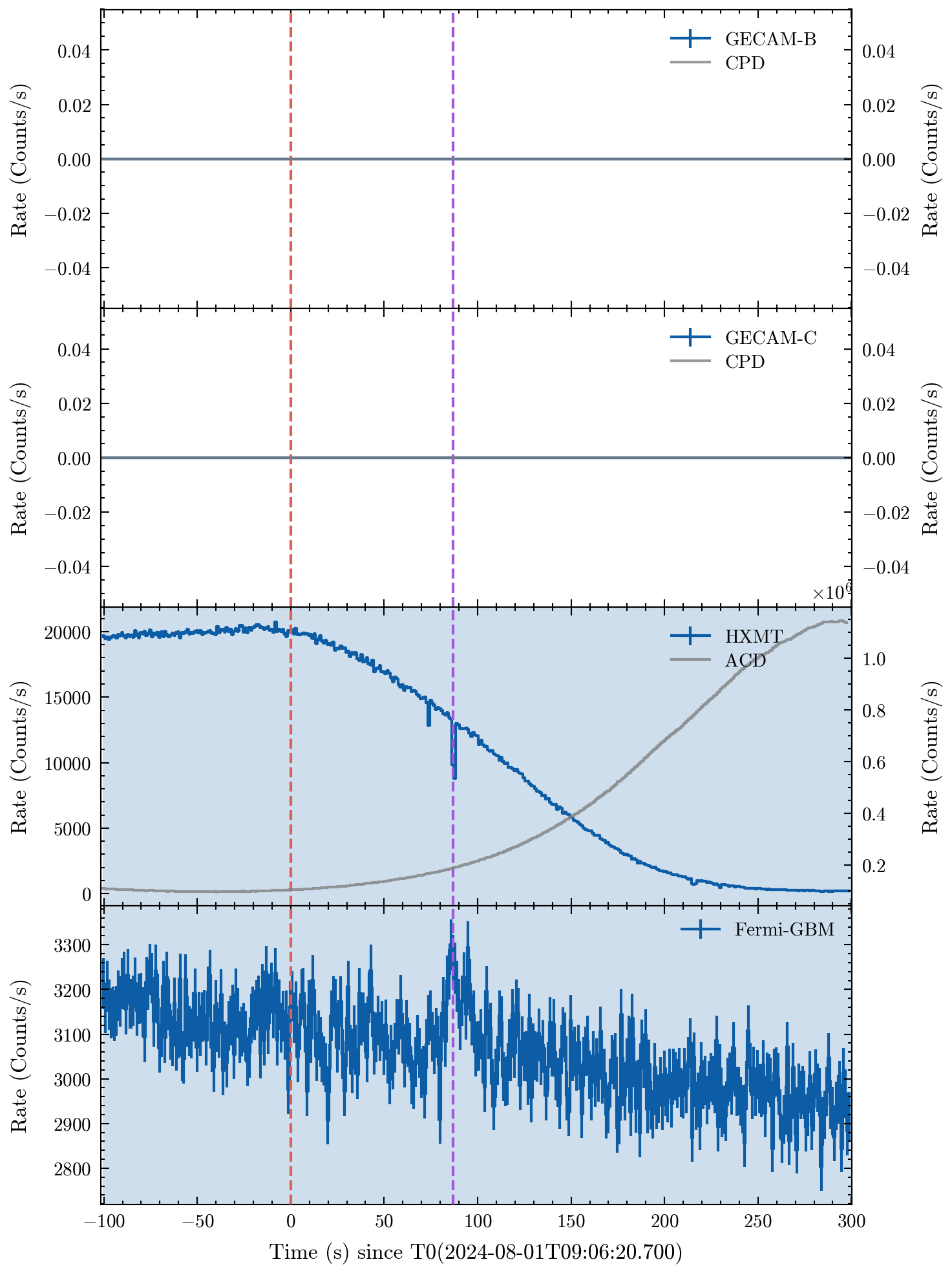}{0.46\textwidth}{(d)}}
\caption{Multiple instrument light curves for 14 EP/WXT X-ray transients whose gamma-ray counterpart is found in this work. Dip in \textit{Insight}-HXMT light curve is caused by a known issue in the time tag of photon events which occasionally occurred in the data. Caption is same as Figure \ref{fig:fitting}. }
\label{fig:search_v1}
\end{figure*}

\begin{figure*}
\gridline{
\fig{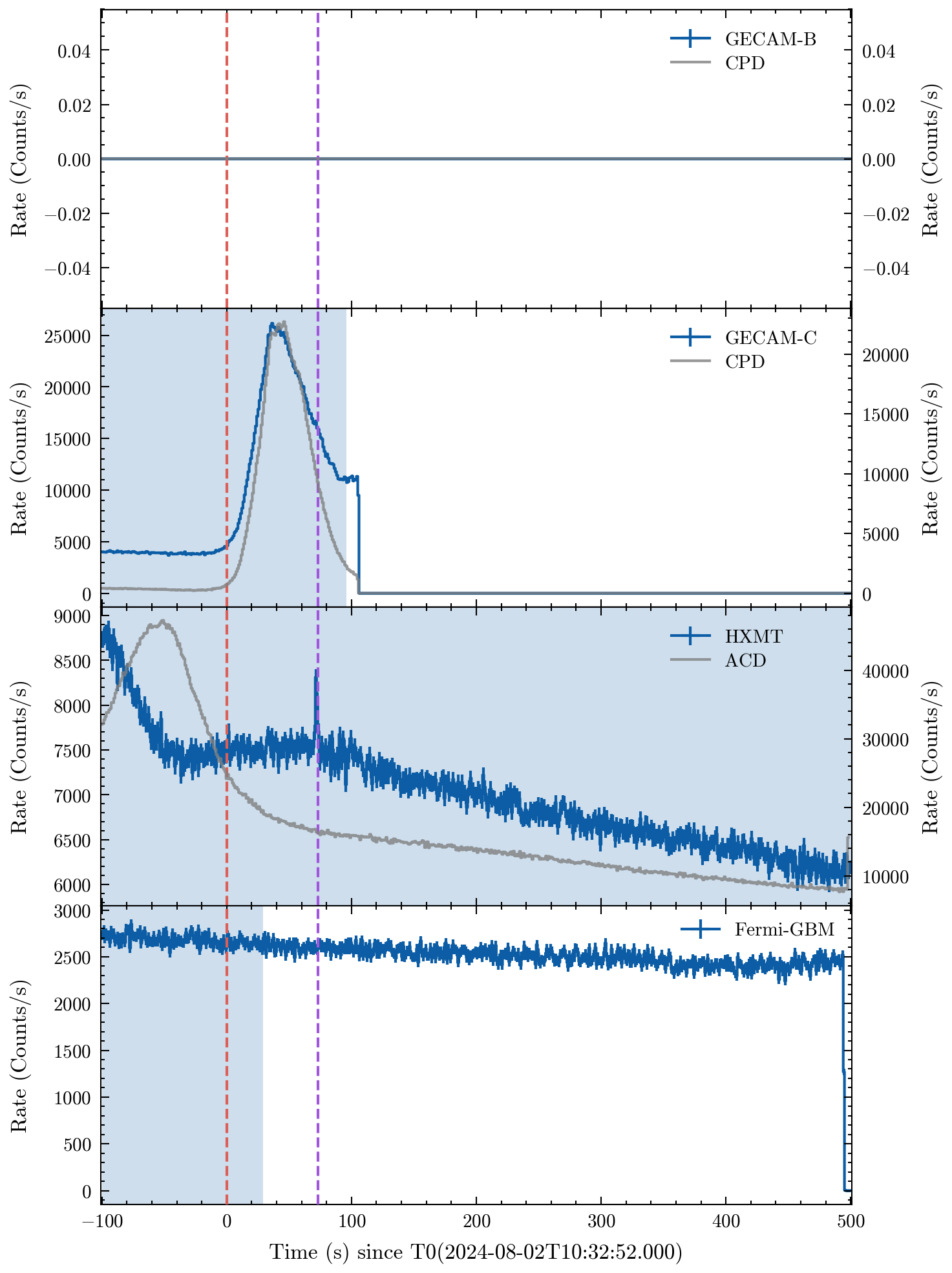}{0.46\textwidth}{(e)}
\fig{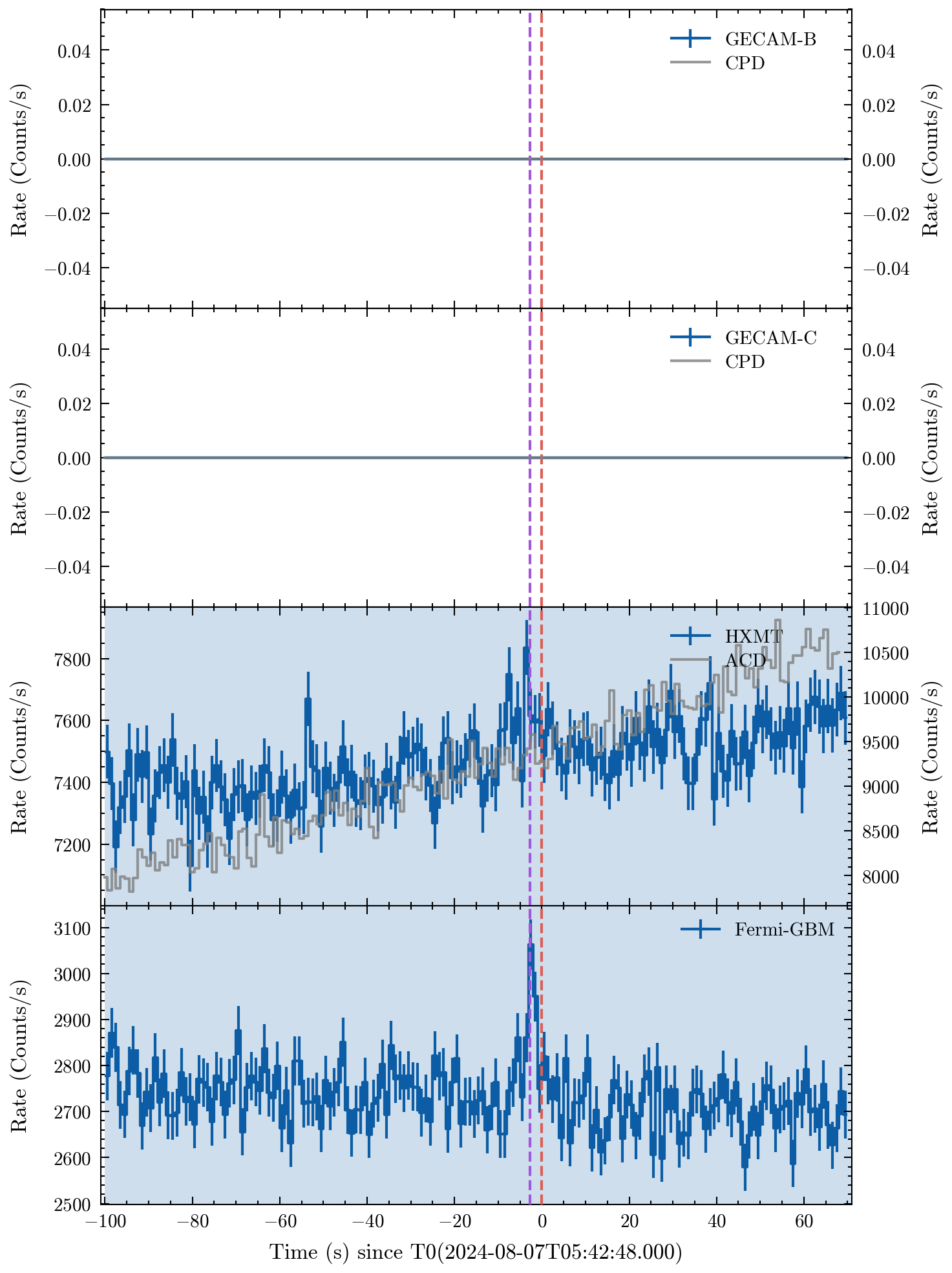}{0.46\textwidth}{(f)}
          }
\gridline{
\fig{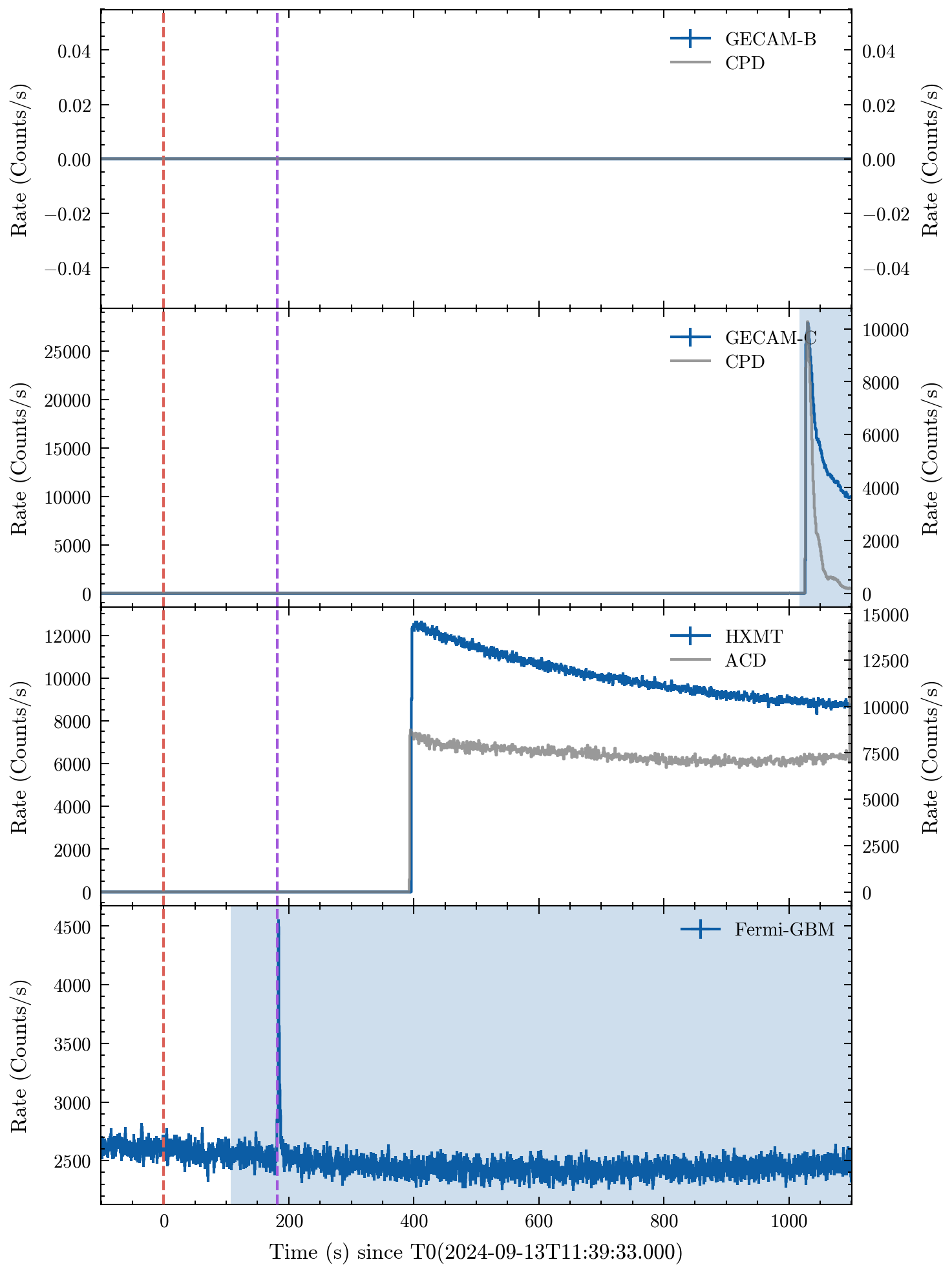}{0.46\textwidth}{(g)}
\fig{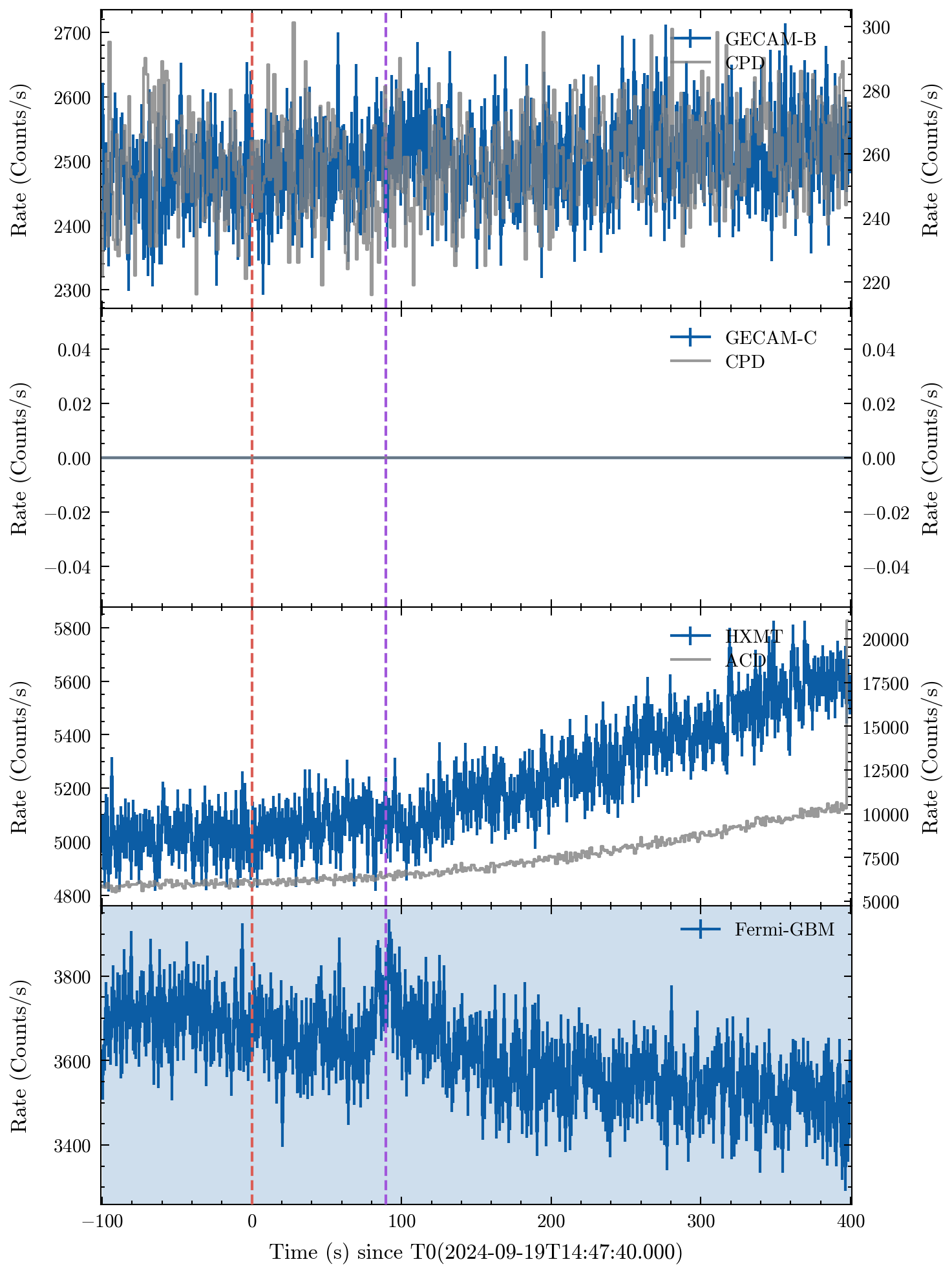}{0.46\textwidth}{(h)}
}
\caption{Figure \ref{fig:search_v1} continued.}
\label{fig:search_v2}
\end{figure*}

\begin{figure*}
\gridline{
\fig{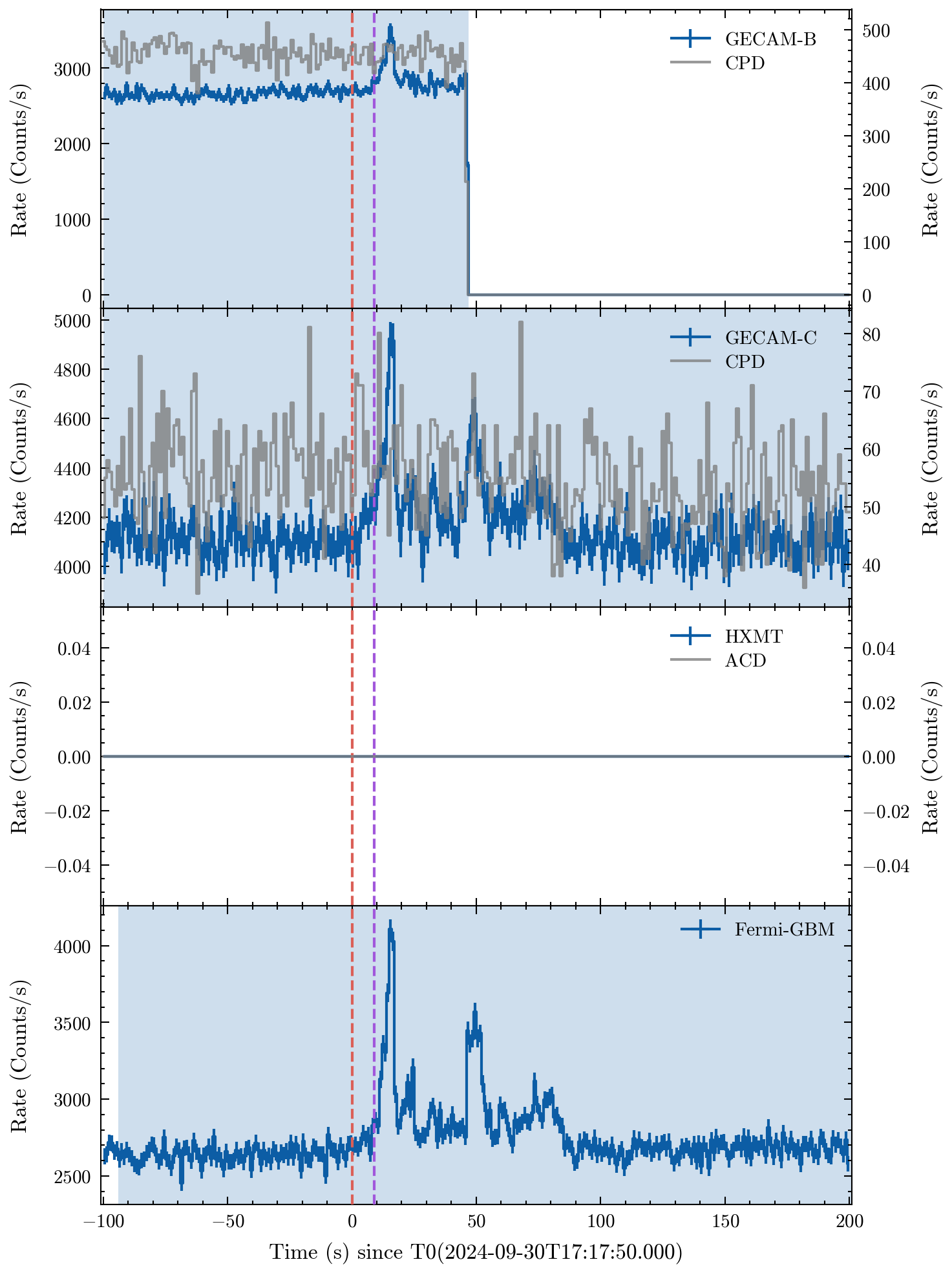}{0.46\textwidth}{(i)}
\fig{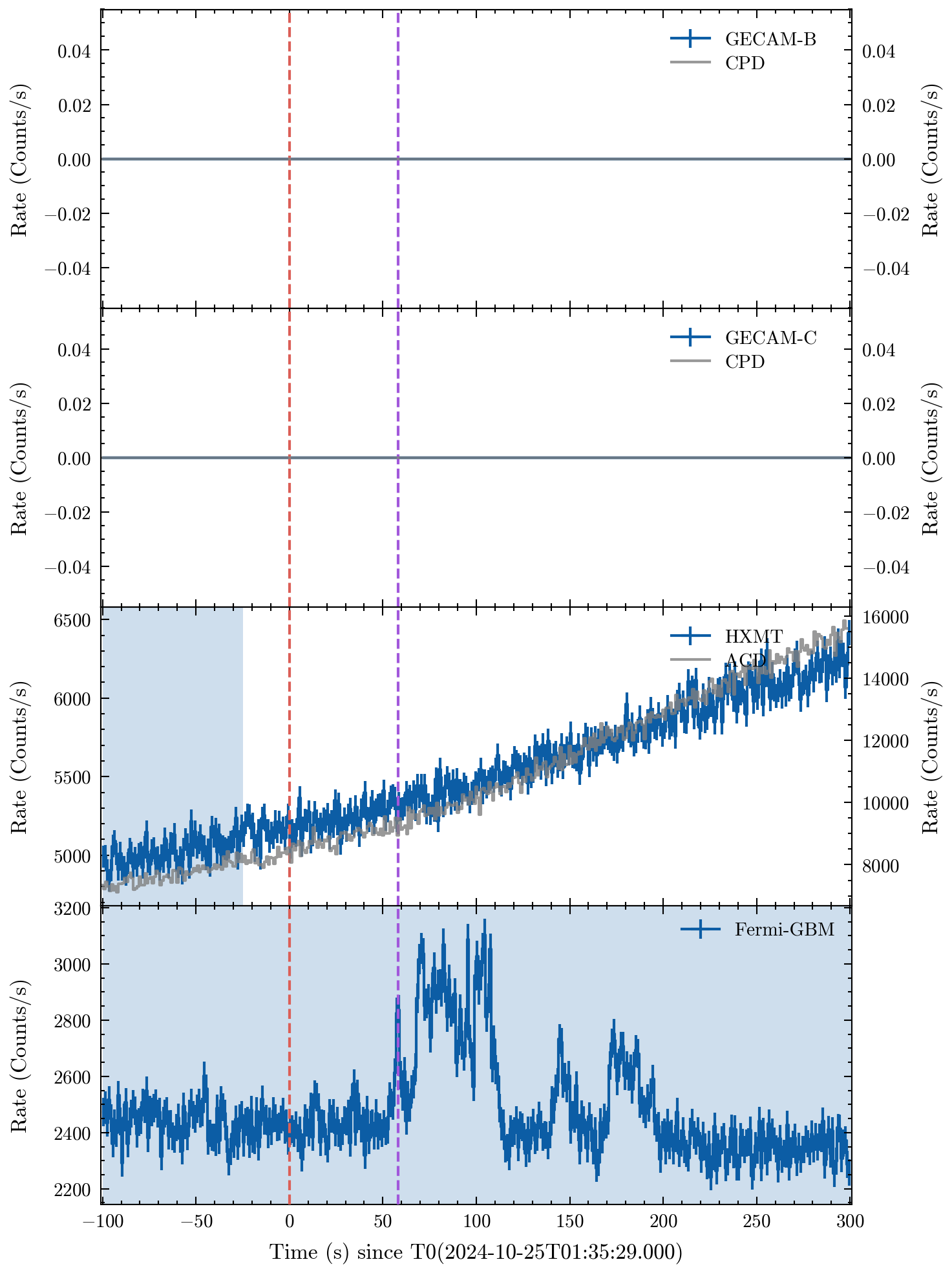}{0.46\textwidth}{(j)}
          }
\gridline{
\fig{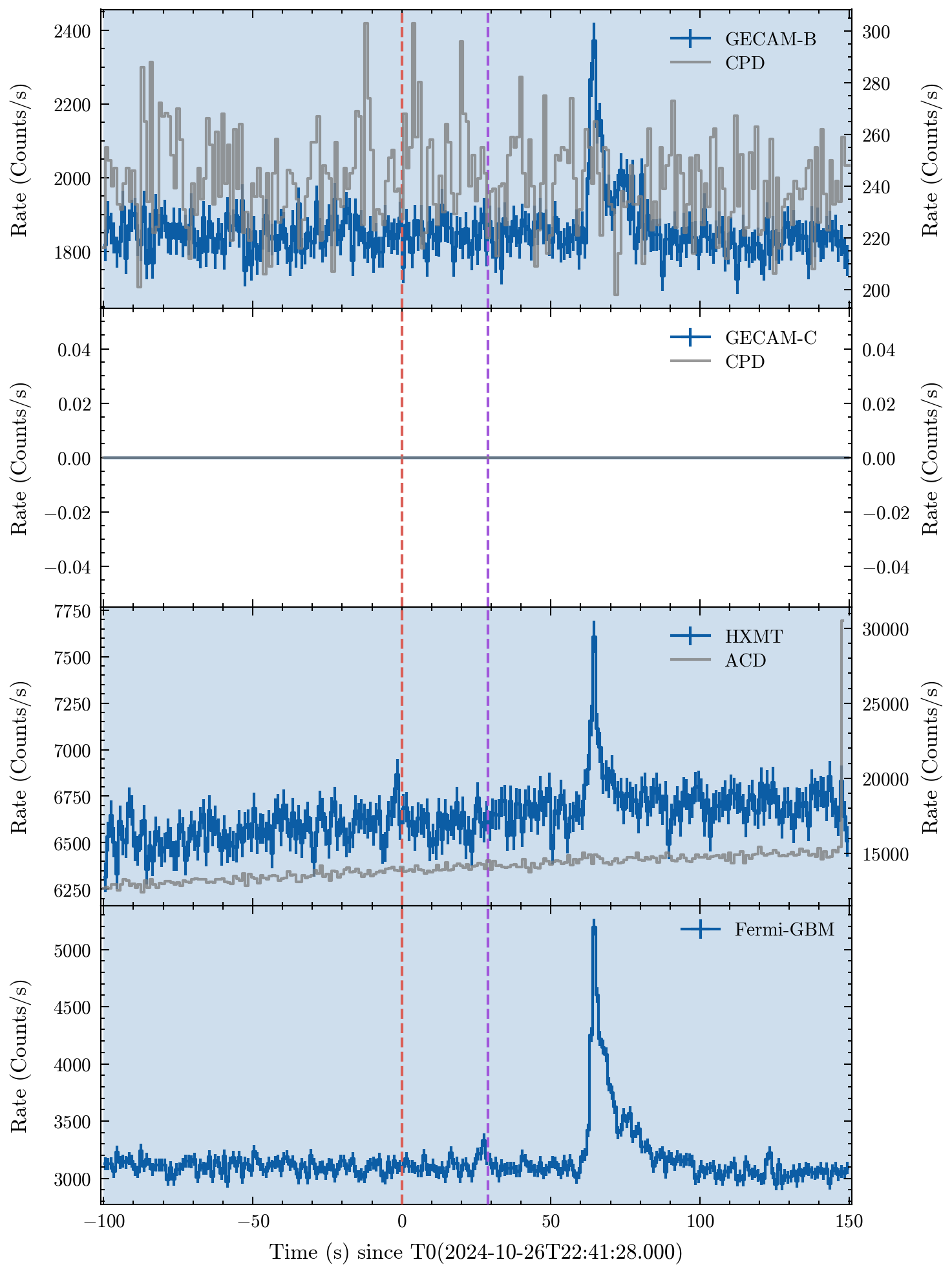}{0.46\textwidth}{(k)}
\fig{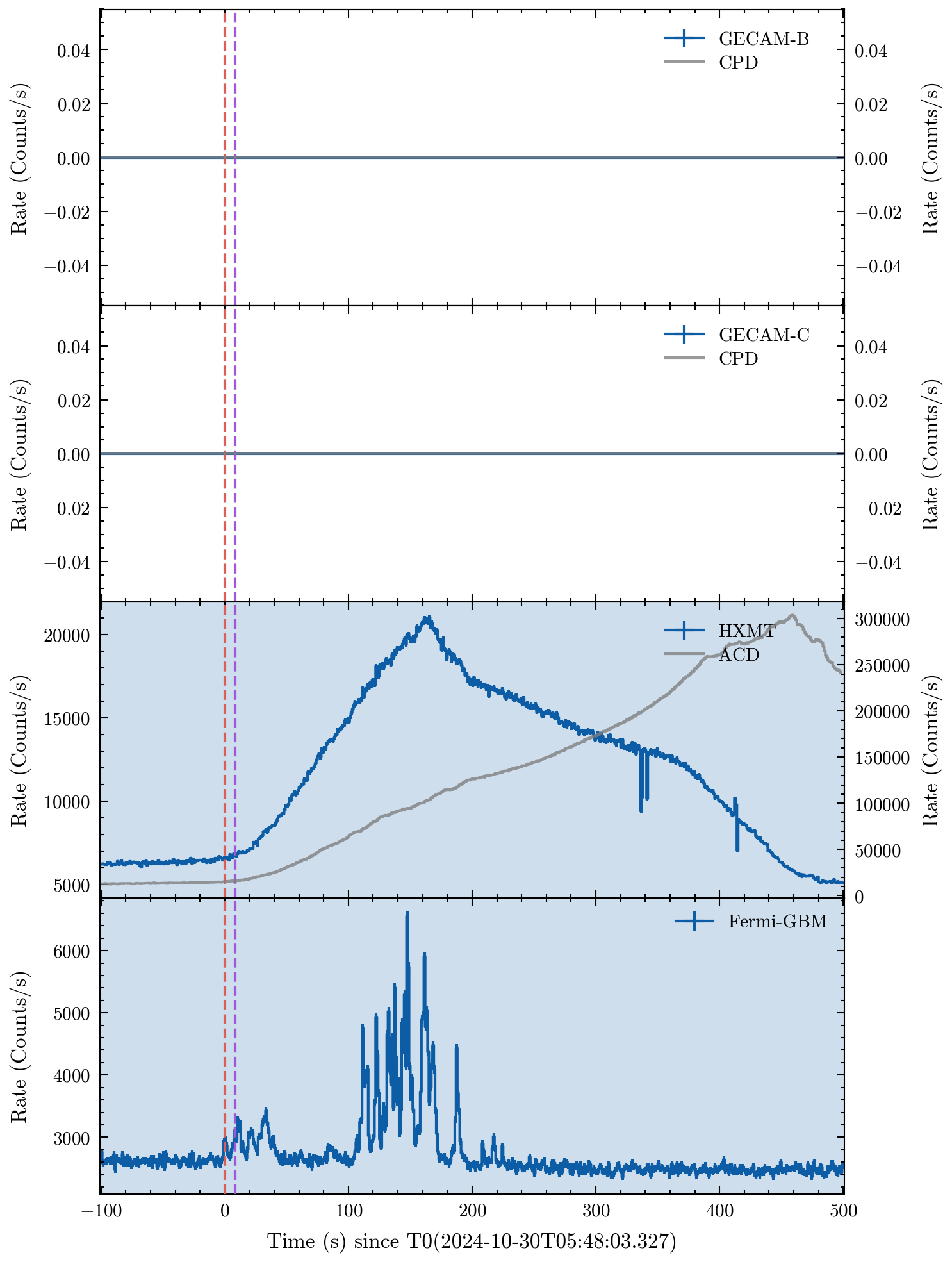}{0.46\textwidth}{(l)}
}
\caption{Figure \ref{fig:search_v1} continued.}
\label{fig:search_v3}
\end{figure*}

\begin{figure*}
\gridline{
\fig{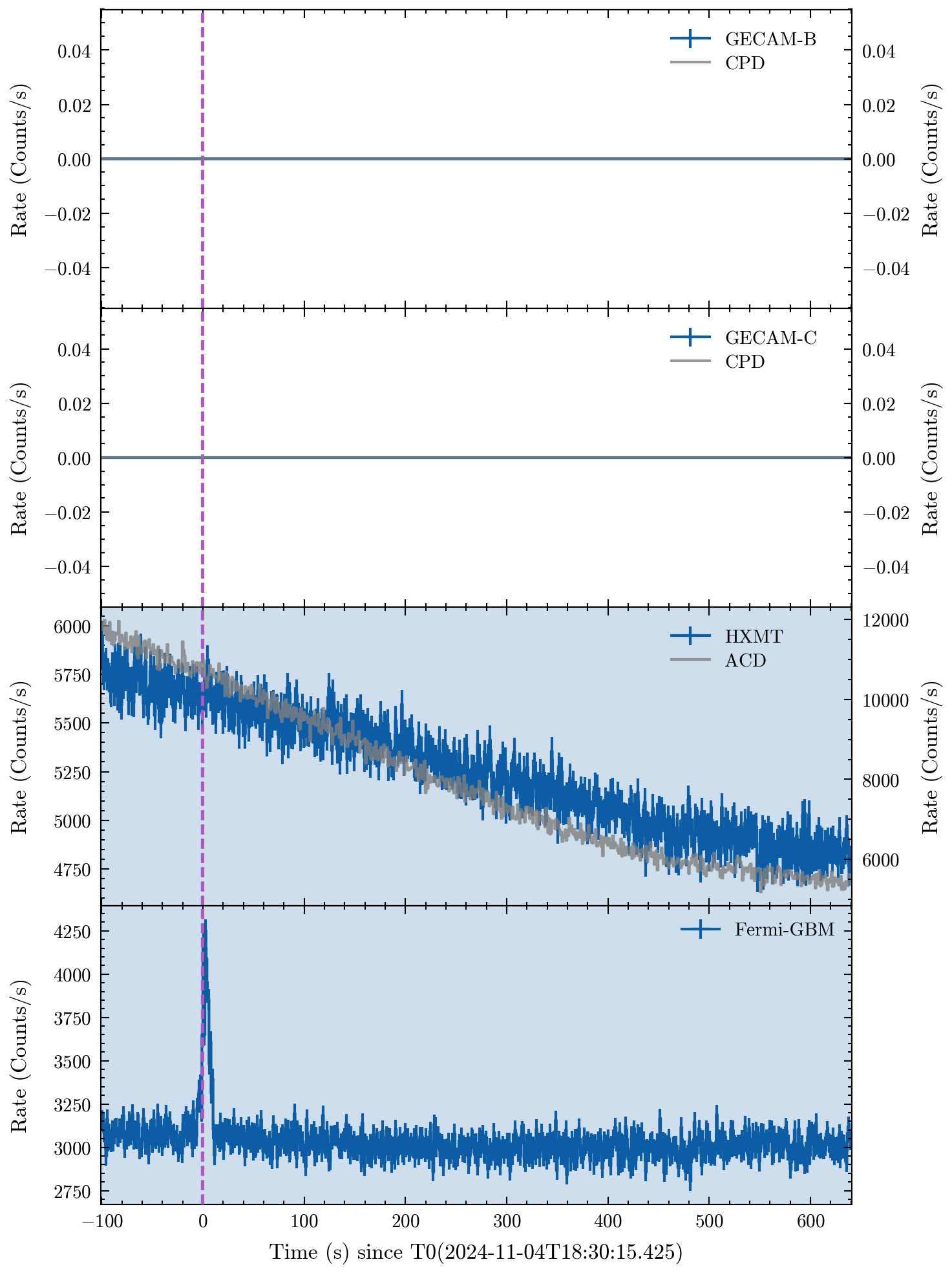}{0.46\textwidth}{(m)}
\fig{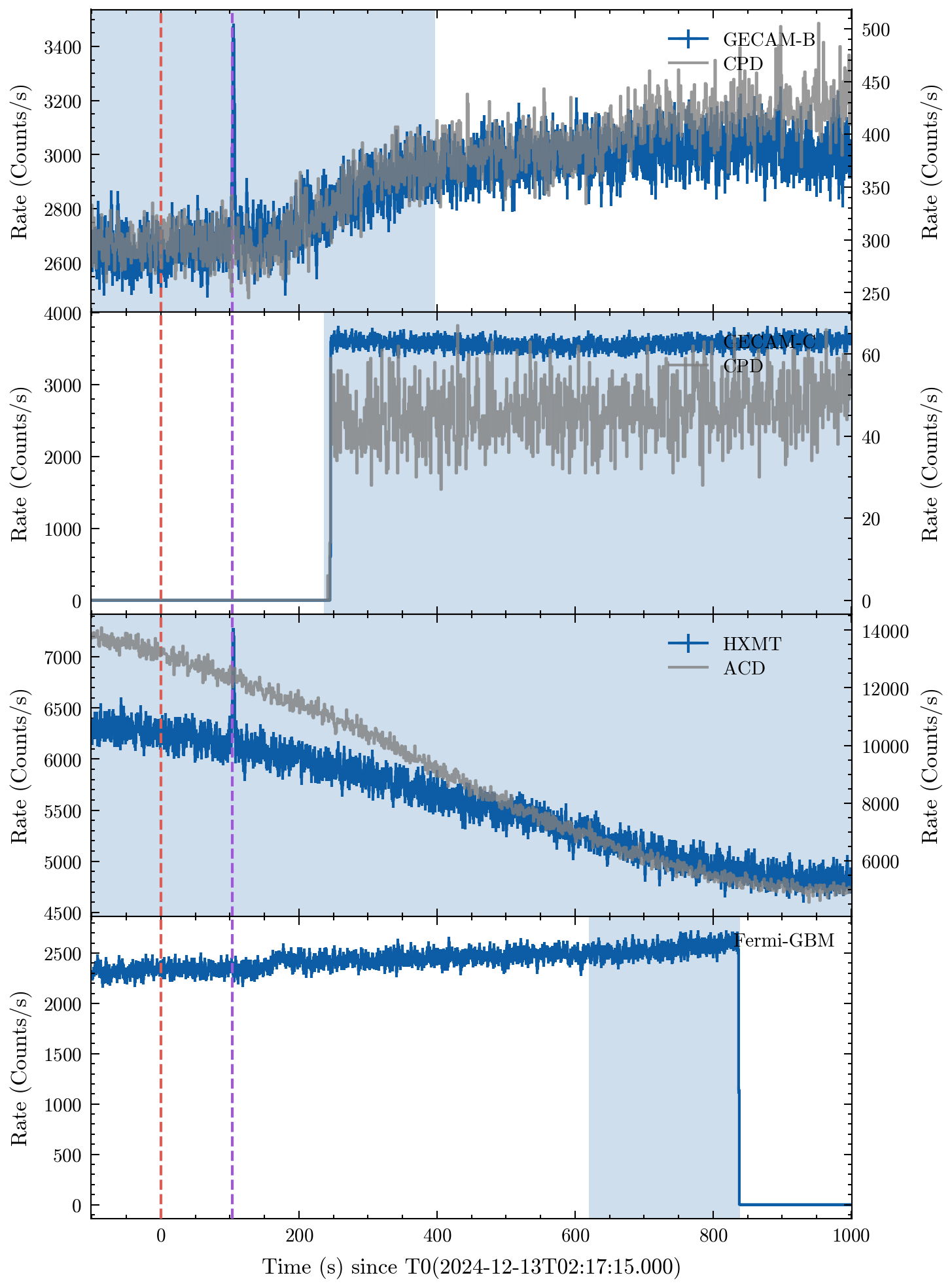}{0.46\textwidth}{(n)}
}
\caption{Figure \ref{fig:search_v1} continued.}
\label{fig:search_v4}
\end{figure*}
\end{document}